%% file: main.tex
\documentclass[Times1COL]{WileyNJDv5}

\usepackage[T1]{fontenc}

\articletype{Article Type}%

\received{Date Month Year}
\revised{Date Month Year}
\accepted{Date Month Year}
\journal{Journal}
\volume{00}
\copyyear{2023}
\startpage{1}

\usepackage{comment}
\usepackage{tabularray}              
\UseTblrLibrary{varwidth}
\usepackage{ragged2e}
\usepackage{enumitem}

\RequirePackage{amsmath}
\RequirePackage{amssymb}

\usepackage[threshold=1,autopunct]{csquotes}

\let\bibfont=\relax
\usepackage[
	citestyle=ama, bibstyle=ama
]{biblatex}
\let\parencite=\supercite
\let\textcite=\supercite
  \let\bibfont\abstractfont

\newtheorem{example}{Example}	
\excludecomment{exampleA}		

\RequirePackage{POE}
\def\POE#{\xdef\POE##{POE}Problem Oriented Engineering (\POE{}~\cite{hall2017a-design})}

\RequirePackage{ebproof}

\newcommand{\POED}{POE-\texorpdfstring{$\Delta$}{D}}
\newcommand\solved{\ifmmode \;\fi\ensuremath{\surd}}
\renewcommand\solved{\relax}
\newcommand{\refinedby}{\triangleright}
\newcommand{\becomes}[0]{\ensuremath{\triangleright}}

\usepackage[capitalise]{cleveref}
\crefname{CPSstep}{step}{steps}
\crefname{clause}{clause}{clauses}
\crefformat{paragraph}{\S{}#2#1#3}
\crefrangeformat{paragraph}{\S\S{}#3#1#4-#5#2#6}
\crefformat{equation}{eq.~#2#1#3}

\newcommand{\currentstate}[0]{\ensuremath{\Sigma}}
\newcommand{\currentproblems}[0]{\ensuremath{\Phi}}
\newcommand{\domadd}[1]{\mathord{+}#1}
\newcommand{\nulldomain}[0]{\ensuremath{\emptyset}}
\newcommand{\update}[0]{\mathbin{\Delta}}
\newcommand{\wrt}[0]{to the satisfaction of}

\input{localDefs.tex}

\begin{document}

\title{\POED{}: A FRAMEWORK FOR CHANGE ENGINEERING}

\author[1]{Georgi Markov}

\author[2]{Jon G.~Hall}

\author[2]{Lucia Rapanotti}

\authormark{Markov \textsc{et al.}}
\titlemark{\POED{}: a framework for change engineering}

\address[1]{\orgname{Siemens Foundational Technologies}, \orgaddress{\country{USA}}}

\address[2]{\orgdiv{School of Computing and Communications}, \orgname{The Open University}, \orgaddress{\country{UK}}}

\corres{Corresponding author Jon G.~Hall, \email{Jon.Hall@open.ac.uk}}

\abstract[Abstract]{Many organisational problems are addressed through systemic change and re-engineering of existing Information Systems rather than radical new design. In the face of widespread IT project failure, devising effective ways to tackle this type of change remains an open challenge. This work discusses the motivation, theoretical foundation, characteristics and evaluation of a novel framework -- referred to as \POED{}, which is rooted in design and engineering and is aimed at providing systematic support for representing, structuring and exploring change problems of a socio-technical nature, including implementing their solutions when they exist. We generalise an existing framework of greenfield design as problem solving for application to change problems. From a theoretical perspective, \POED{} is a \emph{strict} extension to its parent framework, allowing the seamless integration of greenfield and brownfield design to tackle change problems. 

A Design Science Research methodology was applied over a decade to define and evaluate \POED{}, with significant case study research conducted to evaluate the framework in its application to real-world change problems of varying criticality and complexity. The results show that \POED{} exhibits desirable characteristics of a design approach to organisational change and can bring tangible benefits when applied in practice as a holistic and systematic approach to change in socio-technical contexts. 
}



\maketitle

\section{Introduction}
Continuous change is part of the make-up of the modern organisation and crucial to its survival, evolution, differentiation and thus overall success in a business environment of ever-increasing competition, complexity and volatility. 

The main triggers for such change today are very often technology-driven transformations brought about by continuous evolution of legacy systems or increasing adoption and convergence of new and highly disruptive digital technologies\parencite{rezgui2005socio-organisational, tsiavos2021technology}, such as cloud computing, 5G, internet of things, blockchain, or artificial intelligence to name a few. Although technology driven, they will typically trigger far-reaching, transformational changes at many levels of the adopting organisation and beyond, disrupting both technical infrastructures, business models and processes, culture and workforce behaviour\parencite{tsiavos2021technology,nance1996investigation}. Understanding change impact across the entire change context is difficult even for trivial systems\parencite{parashar2011change}, and becomes increasingly so for highly complex systems of systems and software ecosystems due to the additional dependencies (interface, business model, etc.) and complexities involved in orchestrating the cooperative work of the constituent technical systems within their social context.

For example, the shift from DVD rentals to digital delivery for Netflix\parencite{ryan2013leading} and the introduction of music streaming through Apple Music\parencite{moreau2013disruptive}, both enabled by advances in digitisation and broadband network technology, could only be turned into viable business transformations through significant and organisation-wide changes\parencite{halal2015business}, encompassing both technological infrastructure, e.g., new cloud and docker technologies, and micro-service architecture\parencite{leung2017titus}, business models and processes, e.g., the introduction of subscription services\parencite{oat2013analysis} and digital rights management solutions\parencite{montgomerie2013owning}, orgnaisation-wide cultural changes, e.g., the adoption of total quality management\parencite{olson2018total} and agile practices\parencite{lawler2015organization}, and changes in consumers' attitudes and behaviour leading to new market opportunities.

Existing approaches to change are seen as only providing very high-level change process descriptions to guide the organisation through change initiatives. Often focused on soft aspects of management and culture, they tend to be imprecise and unsystematic, creating a disconnect between the technical and the social aspects of change, and lacking support for the analysis of the impact of change on the wider organisational context and its systems\parencite{nielsen2015systems}. This is true\textcite{cao2004need} of even influential approaches adopted in practice, such as Total Quality Management (TQM) and Business Process Reengineering (BPR). Specific discussion points include\parencite{valiris1999critical,reijers2005best}:
\begin{itemize}
\item a lack of support and technical direction when it comes to the re-design of processes and systems  
\item too narrow a focus on descriptions of the \enquote{situation before} and the \enquote{situation after}, giving little consideration to the re-design process itself and its fluidity within an ever changing context.
\end{itemize}

\POED{} was conceived to address these challenges, as an holistic change framework, rooted in design and engineering, but crossing the technical-social divide. \POED{} provides systematic support for representing, structuring, exploring, and solving change problems of a socio-technical nature, including the analysis of change within specific systems and its ripple effects on other interconnected parts within the larger context of the changing organisation, with due consideration of stakeholders' roles and perspectives, and process risk.

\POED{} was developed over a decade of Design Science Research\parencite{peffers2007design}, and validated through significant applications to real-world case studies. 

The work presented here introduces the framework in its entirety, including its theoretical basis, and its formal and informal problem solving processes. Alongside an illustrative example, it also discusses its key features and performance in relation to its application to real-world socio-technical change problems, including an evaluation of its strengths and weaknesses in relation to its use in professional practice. 



\section{Background}
\label{background:sec}

Digital transformation\parencite{gartner2023digital}, be it that of modernising or optimising digital systems via the introduction of new technologies, or inventing new digital business models, is acknowledged as a major trigger for technology-driven organisational change. As a result, several technology-centric approaches have been proposed within the Information Technology (IT) field, revolving around the concept of Enterprise Architecture (EA), a blueprint of the structure and operations of an organisation used to harmonise and standardise processes, information, technology and operation across all sub-systems. Notable examples of EA frameworks adopted in practice include Zachman\footnote{\url{https://www.zachman.com/about-the-zachman-framework}}, ITIL\footnote{\url{https://www.itlibrary.org}}, COBIT\footnote{\url{https://www.isaca.org/resources/cobit}}, and TOGAF\footnote{\url{https://www.opengroup.org/togaf}}. 

These approaches have attracted criticisms for being often impractical and for lacking appropriate implementation guidance\parencite{kotusev2016critical,barroero2010business}. In dealing with business goals, strategies and governance primarily from an information and automation standpoint\textcite{mezzanotte2010applying}, they do not address users' behavioural patterns, so that socio-technical issues are not appropriately accounted for in the change, whether the interaction between people and IT, or the interaction among people mediated by IT\parencite{nielsen2015systems}. Equally, they don't address the wider effects on the organisation, particularly how change should propagate as a result of the various systems dependencies. This is crucial in the context of large Systems of Systems (SOS)\parencite{popper2004system,jansen2019managing}, characterised by complex interrelationships among diverse and multiple systems and stakeholders across organisational units\parencite{knauss2018continuous}. Two additional issues arise in this case: the entities comprising such systems may change at different rates, or may be subject to different regulatory and legislative regimes, possibly  across different countries or even continents\parencite{serebrenik2015challenges}. This brings additional evolutionary challenges, not addressed by current EA frameworks, particularly in the context of safety-critical and mission-critical systems, due to the need to re-gain assurance of key properties when SOS are modified after their initial certification.

In Management Science\textcite{drucker1955management}, organisational change is seen more holistically as \blockcquote{moran2000leading}{the process of continually renewing an organisation's direction, structure, and capabilities to serve the ever-changing needs of external and internal customers.}  As evidenced, for instance, by the comprehensive review\textcite{van-de-ven1995explaining} a vast body of work exists on the topic, which started as far back as the 1950s, with four main kinds of change theory emerging as a result. Briefly, \textit{lifecycle theories} emphasise sequential, well-defined and unavoidable stages in the progression from pre- to post-state. \textit{Evolutionary theories} use the metaphor of natural selection, where changes come from cycles of random variations, followed by best-fit selection, then establishment as the new norm. \textit{Dialectical theories} assume that change is the result of conflict. Finally, \textit{teleological theories} focus on intentional change, where organisations are driven by explicit goals of continuous improvement, optimisation and monitoring. It is worth noting that other classifications exist in the literature\parencite{pawlowsky2001treatment,demers2007organizational,weick1999organizational}, however, while the terminology differs, the underlying concepts are similar to those introduced by\textcite{van-de-ven1995explaining}.

Teleological theories have been the most influential in both the academic literature and professional practice\textcite{kezar2001understanding,uluturk2012assessment}, and have led to the field of Organisational Development\parencite{lewin1947frontiers,chin1969general,havelock1973change,allen2001applying,lunenburg2010organizational}, which focuses on behavioural science techniques to plan and implement change in work settings. While early works are rarely applied in current practice -- as they do not account for today's fast-paced environments, requiring different change phases to co-evolve -- many current approaches to change management stem from those early teleological theories. These include Total Quality Management (TQM) and Business Process Reengineering (BPR\textcite{hill1998positioning}), the former aimed at the continual improvement of the quality of an organisation's products, services, people, processes and environment\parencite{goetsch1995implementing}, and the latter at the radical improvement of organisational processes\parencite{hammer1993reengineering}. More specifically, TQM seeks to integrate all organisational functions (marketing, finance, production, customer service, etc.) to meet customer needs and organisational objectives, taking a holistic approach which involves everyone in the organisation: TQM has influenced quality management standards and methodologies including ISO 9001, Six Sigma, and Lean Manufacturing\parencite{marouni2014certified}. 
 
 On the other hand, BPR considers organisational processes as the main lever for change, advocating radical rethinking and redesign to achieve `dramatic' improvement: it advocates a top-down approach requiring managers to think outside of the constraints of the current organisation (the \emph{as-is} situation) and look at how the work should be performed if the business were to start from scratch (the desired \emph{to-be} situation)\parencite{grobman1999improving,netjes2009bpr}.  

Despite their key difference (\emph{continual} vs \emph{radical change}), both approaches rely on process (re\nobreakdash-)design as an essential step towards improving the status quo. Yet, they are criticised for not providing sufficient guidance as to how this can be achieved in practice\parencite{zand1975theory, lim2014multidimensional, verkerk2004trust,kleiner2000revisiting}. Such criticisms stem for the fact that their focus is on the \emph{why} for the change, rather than \emph{what} should change, hence, offering minimal operational support to practitioners. As a result, such authors advocate the need for a more structured, detailed design theory for change, accompanied by models able to guide the change process in practice.

The idea that organisations are the product of {design} started to develop in the 1970-80s\parencite{galbraith1974organization,simon1977structure,buchanan2008introduction}, with Organisational Design aimed at establishing a scientific basis for the role of design in management and organisational change - often under the term \enquote{innovation}. \textcite{hevner2004design} positions design at the \enquote{confluence of people, organisations and technology}, where both design and behavioural sciences should apply to ensure their strategic alignment while addressing fundamental design problems in technology-driven organisational change. By design problem is meant a real-world problem which is ill-structured and complex\parencite{jonassen2000towards}: ill-structuredness implies an often vague starting point, unclear success criteria and no unique path to solution\parencite{simon1977structure,goel1989motivating}; while complexity refers to the number of variables involved, their stability over time, and their interactions\parencite{funke1991solving,fischer2011process}. This definition applies to any artificial artefact, whether a physical product, a system or just a course of action\parencite{smith1993conceptual}, and has been particularly influential in `greenfield' design, that is when the focus are radically new artefacts. In our research, we strive to extend design problem solving to `brownfield' design, and specifically to organisational change problems. 

The key characteristics of design problem solving\textcite{smith1993conceptual} are as follows: 
Firstly, it must allow the exploration of unsatisfied goals or needs which motivate, inform, and instigate the design activity, and provide evaluation criteria for the designed artefact in its context. Secondly, it must enable the considerations of constraints which arise both from the context and as part of the design activity, and which determine what is feasible within that context. Thirdly, it must allow for design alternatives to be generated and explored through experiential knowledge (personal and/or collective) and creative imagination, alongside reasoned analysis. Design problem descriptions and representations of any form should also be allowed to support the design activity, both individually and in combination and at various degrees of precision, be that pictures, or formal or informal statements, or mental models underlying human thinking: such linguistic diversity is considered essential for communication and sense-making among diverse stakeholders. 

Overall, it is safe to conclude that, despite a sustained effort across many disciplines, an effective approach to organisational change is still lacking, although its desirable characteristics are known. Specifically, that it should allow for:
\begin{itemize}
    \item \textbf{Design problem solving}, exhibiting the key characteristics\textcite{smith1993conceptual}.
    \item  \textbf{Identification of system and sub-systems scope and boundaries}, so that both the overall system and its constituent (sub-)systems are clearly identified, including their interfaces and interdependencies.
    \item \textbf{Diversity of descriptions and perspectives}, in order to capture salient properties of the system,  different stakeholders' perspectives and validation criteria, and various forms of reasoning, in support of system wide analysis, investigation of change impact, and validation of designed artefacts.
    \item \textbf{Factoring assurance within the change process}, in order to support reasoning about the effect of change on key properties of the system, including support for re-certification when needed. 
    \item  \textbf{Process support and guidance}, in order to facilitate the practical implementation of organisational change, including maintaining traceability of intermediate evolutionary steps.
\end{itemize}

In the reminder of this paper we will introduce \POED{}, our candidate framework for organisational change, and argue the extents it exhibits thesse characteristics based on its empirical evaluation to date.

\section{Methodology}\label{sect:methodology}

\POED{} was defined and evaluated over a decade of Design Science Research\parencite{peffers2007design}, following an iterative process in which a range of change problems with diverse characteristics were studied to inform and validate the framework design and its performance in real-world application. The studies are summarised in Table~\ref{studies:t}, details in\parencite{markov2024poe-delta}. In Section~\ref{sect:evaluation}, we will discuss outcomes and main findings as part of the overall evaluation of the framework.

\NewColumnType{R}[1][]{X[#1,preto={\RaggedRight}]}
\begin{table}[htbp]
\footnotesize
\centering
\begin{tblr}{
  colspec={c|RlRRRRR}, 
  hline{2}={0.5pt},
  row{1}={b,font=\bfseries},
  colsep={2pt},
}
Study & Change problem & {Scale/Criticality\\/Duration} & Context and stakeholders & Scope of Change & Evaluation Aim & Primary Findings & Empirical Evidence \\
\hline
NICE &
Re-design of online presence, including integration with social networks and online member registrations &
{Small-scale\\Low criticality\\3 months} &
Private non-profit group of English-speaking families in Germany; participants included website administrators, members, and end-users &
Brownfield: Enhancing existing digital systems &
Establish general applicability, usefulness, usability, and needed improvements &
Highlighted ease of initial problem structuring and stakeholder engagement; emphasised need for tool support &
Positive user comments on usability and usefulness; stakeholders appreciated structured problem-solving. \\
OSLC &
Integration of various engineering tools across multiple international organisations &
{Large-scale\\High criticality\\9 months}&
Multinational engineering company; involved software engineers, project managers, and integration specialists from different organisations &
Brownfield: System integration across organisational boundaries &
Evaluate applicability and effectiveness for complex change within interdependent systems &
Demonstrated ability to handle tangles and cross-organisational dependencies; suggested need for tool-support and automation to help managing the many and complex interdependencies between domains &
Observed significant improvements in system integration timelines and inter-team communication; practitioners noted increased problem clarity \\
ATC &
Forensic analysis of a historic project to generate reusable insights and inform the design of a highly automated train dispatch system &
{Medium-scale\\High criticality\\4 months} &
Multinational engineering company; participants included systems engineers, project managers, and process analysts &
Brownfield: Retrospective analysis of historic system changes for reuse in new system &
Evaluate forensic capabilities for capturing and analysing historic changes, and reusing solutions &
Revealed potential for forensic analysis to guide new projects; validated design traceability features &
Identified reusable design patterns and process inefficiencies; stakeholders reported improved understanding of past failures \\
LLM-A &
Transition from manual troubleshooting to an LLM-driven system &
{Medium-scale\\Medium criticality\\Phase 1: 6 months\\ Phase 2: ongoing} &
Building automation company managing automation in industrial buildings; participants included AI researchers, field technicians and system architects &
Hybrid - Greenfield / Brownfield: Legacy system decommissioning and AI design \& integration &
Test end-to-end process and greenfield/brownfield integration &
Showed iterative refinement between \POED{} and \POE{}; highlighted challenges in integrating socio-technical changes and addressing emergent requirements &
Participants confirmed \POED{}'s ability to align greenfield and brownfield processes dynamically and noted iterative validation as a strength in dealing with emergent challenges. \\
{AI-A\\(Kettle)} &
Replacement of kettle controller from mechanical to AI-driven &
{Small-scale\\High criticality\\N/A} &
N/A: illustrative example only &
Brownfield: Replacing a safety-critical controller &
Validate ability to reason about safety-critical change using a mix of formal and informal notations &
Demonstrated ability to support rigorous reasoning for safety assurance in the presence of linguistic diversity &
Formal simulation confirmed the consistency of formal and informal reasoning and validation of safety-critical decision pathways. \\
\end{tblr}
\caption{Studies conducted to develop and evaluate \POED{}}
\label{studies:t}
\end{table}

\section{The \POED{} Framework}
\label{poed:sec}


The \POED{} framework provides systematic support for solving (representing, structuring,  exploring, forming implementable solutions to) socio-technical change problems. To do so, we generalise the existing \POE{} framework for greenfield problem-solving.

\subsection{Philosophical basis}

Our framework is phenomena centric, by which we mean that its elements are defined in terms of phenomena, with behaviours defined in terms of observable phenomena occurrences and needs in terms of posited phenomena relationships. 

We accept that a focus on observable phenomena can lead to an incomplete understanding of systems, their underlying structures, causes, and of theoretical underpinnings that are not directly observable. We argue, however, that our approach is practical, adequate, and effective in the context of engineering and change management where working with modular system views is a mean by which complexity  an be managed, and that the framework is intended for use by experienced practitioners who bring their own understanding. Our case studies demonstrate that a phenomenon-focus provides effective change problem solving for resource-bounded socio-technical change by allowing:
\begin{description}
\item [Implementing with validation:] by dealing with observable phenomena, engineers can implement changes and validate their effects reliably;
\item [Inferring underlying mechanisms when necessary:] through careful analysis of the requirements for change, underlying structures are explored on an as-needed basis;
\item [Managing complexity:] it permits abstraction and modularity when dealing with change;
\item [Adapting theoretical models:] the use of phenomena does not exclude the use of theory, although, it does ensure that theory use is grounded in observable reality.
\end{description}
Thus, our phenomena-centric focus provides a balance between practice and theory simultaneously acknowledging the importance of underlying mechanism\parencite{hall2017a-phenomenal} while focusing on the aspects of the system that can be directly engaged with for effective change.

\subsection{Conceptual underpinnings}\label{conceptualunderpinnings:s}

\POED{} was developed from \POE{} over a decade of case study research. Like \POE{}, \POED{} offers both \emph{micro-} and \emph{macro-}processes for problem solving; unlike \POE{}, these are for the brownfield problem setting.



The underpinning concepts that \POED{} shares with \POE{} are:
\begin{itemize}
\item systems are characterised as collections of tangled solutions, their future as unsolved problems;
\item stakeholder-centric notions of fitness-for-purpose are used, 
	see~\Cref{stake:phen:s};
\item the micro-process is defined in terms of transformations that relate a single problem and its solution to other problems and their solutions; see~\Cref{micro:s};
\item the macro-process provides a workflow for change, managing the risks involved in problem solving in volatile and complex socio-technical situations; see~\Cref{macro:s};
\item in theory, each step in the macro-process is reducible to, typically, many steps in the micro-process -- whether this must be done is a function of the need for precision.
\end{itemize}

\subsubsection{Problem orientation}

Both \POE{} and \POED{} are \emph{problem} oriented. Problem orientation conceives of the world in terms of problems, where a problem is \enquote{a stakeholder's recognised need in context}\parencite{rogers1983nature}. There are two distinguished problem oriented viewpoints: 
\begin{itemize}
\item that of the \emph{problem owner}, who \emph{recognises} the  \enquote{need in context}; and 

\item that of the \emph{problem solver}, who must understand the problem owner's need and deliver a solution to their satisfaction. 
\end{itemize}

\POE{} does this only in the greenfield setting, i.e., when the solution is a new artefact, to be inserted into a known context\textcite{hall2007arguing,hall2008problem,hall2012software,nkwocha2013design,costantini2021using}\footnote{Known, but not necessarily fixed, contextual forces can change the context during problem solving.}. 

\POED{} widens the notion of problem from greenfield to \textit{brownfield} engineering. In brownfield engineering, the context of the problem is changed as part of the problem solving process. This is a more complex setting; see~\Cref{brownfield:s} for further discussion.
%

\POE{} and \POED{} differ in the form of a problem, that of \POED{} being a generalisation to the brownfield setting of that of \POE{} which was greenfield specific; see~\Cref{changeEng:ss} for the relationship. In the sequel, we focus on the \POED{} framework, noting the relationship with \POE{} only when necessary.

\subsubsection{The challenges of brownfield problem solving}\label{brownfield:s}

Brownfield problem solving is a much more involved process than the greenfield case.

The greenfield setting is defined such that no contextual domain can be changed as part of process of creating the solution(s), which is expected to interact with its environment through phenomena that already exist therein, with the solution being able only to refer to or assert control over them\parencite{jackson2001problem}. One ramification is that implementation of a solution is simply the solution's inserting into the problem environment\footnote{In reality, there may be issues with implementation, but these will be issues caused by considering an in actuality brownfield change as greenfield.}.
 
In contrast, in the brownfield setting the solution will typically involve the creation of new artefacts together with changes to the environment. For this to be effective in discharging the need, the contextual changes must be consistent with other needs of the organisation, such as being consistent with already existing systems on which the change impinges, as must the process by which the new solution is implemented. This complicates considerably the ways in which brownfield change can be obtained: a critical part of a change solution is the development of implementation -- essentially, a project plan -- by which the change will be effected; see~\Cref{micro:s} for a discussion. Neither can the change nor the project plan stand alone as solutions, they must be created and work together synergistically for stakeholder satisfaction.

\subsubsection{Phenomena}\label{phenomena:s}

Like \POE{}, \POED{} uses Jackson's concept of \emph{phenomena} as the basis of problem descriptions. 
According to Jackson\footnote{Original emphasis.}, \blockcquote{jackson2001problem}{[A phenomenon is] an element of what we can observe in the world. Phenomena may be \emph{individuals} or \emph{relations}. Individuals are \emph{entities}, \emph{events}, or \emph{values}. Relations are \emph{roles}, \emph{states}, or \emph{truths}.}

Descriptions of problem components -- the \emph{environment}, the \emph{need}, and the \emph{solution} -- are then descriptions of the relations that exist between those phenomena.

%
%
%
\subsubsection{Modularity and domains}\label{stake:phen:s}

\newcommand{\Gphen}[1]{\rho(#1)}
\newcommand{\Dphen}[2]{\rho_{#1}(#2)}

%
%
For modularity, related phenomena\footnote{By a stakeholder; two stakeholders may see different domains.} might be collected into \emph{domains} by a stakeholder. Two stakeholders may, however, identify different domains and have different understandings of the relationships between phenomena. 
    
Behaviour is captured as relationships between phenomena occurrences with a domain's behaviour being the obvious extension. There is no mandated language for the expression of behaviour, but it is likely that a causal semantics similar to, but not necessarily precisely that of\textcite{moffett1996model} will be needed.

\begin{exampleA}
Gerry, an office worker, is tasked to review policy documents issued by the policy office once per month and to create an update agenda for those that need updating in the light of recent legislation. 

Here, Gerry is a stakeholder and their phenomena of interest include the policy documents and any recent legislation, which they observe, and the agenda that they populate for update, which they control. 
\end{exampleA}


%

\subsection{The \POED{} model of the organisation}\label{PODorgmodel:s}

Organisations are diverse, ranging from small to large, monolithic to multi-part, end-oriented from service to retail, operating within a more or less complex, volatile, and hostile context that can be local or international, located within a supply chain, with various legal and regulatory requirements, and subject to attacks from negative agents.

In this paper, an organisation is seen as (a more or less complex) socio-technical system: a collection of human and technological elements in a more or less fluid structure that interact to determine its behaviour. 

These internal and external characteristics determine an organisation's \emph{problem space} -- the problems it faces in operation that provide motivation for change -- and its \emph{solution space} -- the social and technical areas in which solutions to those problems will be found, as well as the constraints that limit the choices available to it.

Given this we define an organisation to be a pair:
\begin{equation}
	(\currentstate, \currentproblems)\label{orgState:eqn}
\end{equation}
where
\begin{itemize}
\item[\currentstate] the organisation's \emph{current state}, is determined by the collection 
	of interacting solutions to problems the organisation has previously faced, solved, and subsequently embedded
\item [\currentproblems] the \emph{current problems}, determined by the collection 
 	of needs that the organisation currently faces that are, as yet, without solution within the organisation.
\end{itemize}

\subsubsection{Organisational change scenarios}

An \emph{organisational change scenario} (briefly, \emph{scenario}) consists of a change owner, $G$, whose wish it is to satisfy a need $N\in \currentproblems$ in the organisational context $\currentstate$, written
\begin{equation}
	(\currentstate,N)\label{orgScenario:eqn}	
\end{equation}
where the $\currentstate$ and $N$ are as understood by $G$.


Although we identify a change owner whose initial understanding is used in~\Cref{orgScenario:eqn}, we do not assume that this is the definitive view, or indeed that any single individual holds a definitive view. Moreover, no two individuals are assumed to necessarily agree on any detail of either. 

Neither do we assume that the organisation's context nor its needs remain stable during problem solving: both may change over any time period. What is more, we do not assume any individual involved in the organisation is necessarily aware of those changes. This raises substantial risks, of addressing an outdated need, for instance; the macro-process of~\Cref{macro:s} has explicit validation points at which such risks can be managed.

The structures by which an organisation changes are as varied as organisations themselves. We assume that, for substantial changes, change problems owners will delegate the task of solving a change problem to others. In particular, there is likely to be a delegatory chain of responsibility between change parties:
\begin{description}
	\item [change problem owner:] Initiates the change by identifying the need and delegates the problem-solving task;
	\item [problem-solving delegate:] Analyses the problem, explores solutions, and develops an implementation plan for validation by the Problem Owner, but may not implement the solution directly;
	\item [implementation delegate:] Executes the implementation plan devised by the problem-solving delegate and validated by the Problem Owner, bringing the proposed changes into effect.
\end{description}
%

Finally, we note that there is nothing in the definition that prevents both $\currentstate$ and/or $\currentproblems$ from changing during problem solving, so that both $\currentstate$ and $\currentproblems$ are unlikely to be simple (untimed) sets.

As in \POE{}, these organisational characteristics raise problem solving risk, which must be managed \wrt{} the various stakeholders. This is again done in the macro-process.

\section{Engineering change in \POED{}} 

Given an organisational scenario, $(\currentstate,N)$ (\Cref{orgScenario:eqn}) we define an organisational change problem\footnote{In the sequel, simply \emph{change problem} or \textit{problem}} as:
\begin{equation}\label{POED:eq}
\currentstate\update F\meets{G} N	
\end{equation}
which should be read\footnote{We often omit the phrase \enquote{$G$-validated} from environment and need when clear from context.}
\blockquote{
	Starting from the existing 
	environment $\currentstate$, find implementable changes $F$ such that the updated environment $\currentstate\update F$ meets the 
	 	need $N$ \wrt{} problem owner $G$}

Thus, a problem is solved only when $F$ is a complete description of an implementable change that establishes an organisational state satisfying need $N$ \wrt{} the stakeholder $G$.

When implemented, a $G$-validated solution $F$ to problem $(\currentstate,N)$ will change the organisation from $(\currentstate,\currentproblems)$ to 
%
 \begin{equation}
(\currentstate\update F,\currentproblems\setminus\{N\}) \label{orgchange:e}
 \end{equation}
so that the satisfied need is removed from the current problems.

%

\begin{exampleA}[The phone upgrade problem]\label{phoneupgrade:ex}
The \emph{phone upgrade problem} occurs within various contexts:
\begin{itemize}
\item phone: the phone to be replaced
\item financial: whether sufficient funds are available to afford the upgrade
\item technological: whether an data-transfer path exists to the replacement phone
\item physical: whether current accoutrements will make the transfer to the new phone
\item recycling: whether and how the old phone can be recycled
\item time: whether the window for purchasing and receiving the new phone fits within availability
\end{itemize}
We can see these as implementation issues: each will need to be satisfied (or satisficed) for a solution to the upgrade problem to exist. In addition, but tangled within, the choice of which precisely which phone make and model to choose is an additional problem. 

The environment consists, then, of various phenomena which might be arranged in terms of an environment consisting of $[Funds, Tech, Physical, Recycle,Diary]$ giving
\[[Phone, Funds, Tech, Physical, Recycle,Diary]\update F \meets{G} New\_Phone\]
\end{exampleA}

We now turn to the formal and informal processes by which a solution will be found.

\subsection{The micro-process}
\label{micro:s}

The micro process within the \POED{} framework serves as the detailed, step-by-step approach to solving change problems, facilitating precise navigation through complex socio-technical landscapes. It is structured as a Gentzen-style sequent calculus, enabling systematic manipulation and transformation of change problems, represented as expressions that relate existing environments, proposed changes, and the needs they aim to fulfil. Each step in the micro-process links problem-solving activities to concrete, implementable actions, ensuring that the progression from problem identification to solution adheres to rigorous analytical standards. This approach underscores the importance of maintaining stakeholder validation and managing risk while exploring alternative solutions and ensuring consistency with broader organisational goals. 

\subsubsection{Change problem description in \POED{}}\label{changeexpr:s}

\POED{} uses a formal Genzten-style sequent calculus\parencite{kleene1964introduction} for the detailed engineering of change. In theory, all change problem solving steps must be reducible to combinations of applications of rules defined in this section although, whether in practice, this actually happens is a function of the criticality of the problem solving context: for instance, safety- or mission-critical organisational change will tend to micro-process for problem solving while less or non-critical changes will tend to the macro-process.



A \emph{change expression} is defined by the following grammar:
\begin{definition}[Change expression]
\label{POE-Delta-Grammar:defn}
\begin{align}
	NewDomain &:= \domadd{Domain}\label[clause]{adddom:e}\\
	RefineDomain &:= Domain \refinedby \delta[Change](\{NewDomain\})\label[clause]{refinedom:e}\\
	CancelDomain &:= \cancel{Domain}\label[clause]{cancel:e}\\
	Changeable &:= 
						CancelDomain \mid RefineDomain\seq Changeable\label[clause]{changeable:e}\\
	Domain, NeedDesc 
				&:=
						Name: Description\label[clause]{dom:e}\\
	Change &:= \{Changeable\}\label[clause]{change:e}\\
	Environment &:= \{Domain\}\label[clause]{env:e}\\
	Need &:= \{NeedDesc\}\mid Need \parallel Need \mid Need \seq Need\label[clause]{need:e}\\
	ProblemOwner &:= Name[clause]\label{po:e}\\
	ChangeExpression &:= Environment\update Change\label[clause]{chgexp:e}\\
 Problem &:= ChangeExpression \meets{ProblemOwner}Need\label[clause]{problem:e}
\end{align}
where $\{Item\}$ represents a list of $Item$s.
\end{definition}

The grammar is described through the rules define in the next section.

\subsubsection{Change problem transformations}

Change problems transformations related change problems. The transformations defined relate to the formal grammar above.


\begin{exampleA}[Author's problem]
An example of a change expression for the author's problem above is the following expression:
\[\begin{array}{l}
[Manuscript, Audience, Publisher, Formatter]\\
\qquad\qquad\update Manuscript[TextBeforeError,TextAfterError](ReplacementText)
\end{array}
\]
where we identify the domain to change ($Manuscript$, whose description includes the domains that follow) together with its subdomains of correct $TextBeforeError$, the $ErrorText$ replaced by $ReplacementText$, and the $TextAfterError$. The actual replacement text (its description) will typically be defined elsewhere, in a data dictionary, for instance. Domains not referenced in the change expression remain unchanged.
\end{exampleA}

\POED{}'s transformations extend those of \POE{}, and so we use the same Gentzen-style sequent calculus formulation. The general rule is:
\newcommand{\using}[2][]{$\substack{\normalfont \textsf{#1}\hfill\\[0.2ex]\textrm{<#2>}}$}
\begin{equation}
	\begin{prooftree}
			\hypo{E_1\update F_1\meets{G_1} N_1}
			\hypo{...} 
			\hypo{E_n\update F_n\meets{G_n} N_n}
		\infer3[\using{J,I}]{E\update F\meets{G} N}	\label{ruleform:e}
\end{prooftree}
\end{equation}
which should be read as:
\blockquote{$F$ is a solution to change problem $(E,N)$ \wrt{} $G$ if $F_i$ is a solution to $(E_i,N_i)$ \wrt{} $G_i$, and $J$ (resp. $I$) extends the justification $J_i$ (resp.~implementation $I_i$) of $F_i$ \wrt{} $G$.}


\paragraph{Delegation}\label{delegate:p} We illustrate the general rule form through an example.

The act of delegation between problem owner and trusted problem solver as a problem transformation is:
\[\begin{prooftree}
		\hypo{E_D\update F_D\meets{D} N_D}
	\infer1[\using[Delegation]{$G$ trusts $J_D$, $G$ trusts $I_D$}]%
		{E_G\update F_D\meets{G} N_G}
\end{prooftree}\]
which should be read as
\blockquote{$F_D$ is a solution to change problem $(E_G,N_G)$ to the satisfaction of  $G$ if $F_D$ is a solution to $(E_D,N_D)$ \wrt{} $D$, $G$ trusts their delegate's justification $J_D$ and implementation $I_D$.}
or the delegate's solution becomes the problem owner's solution when it satisfies the delegate, and $G$ is satisfied by the delegate's reasoning.

From this example, we see how the relationships between stakeholders drives the problem solving micro-process.For rule class definitions, we will add a name the rule, here \enquote{\textsf{Delegation}}, for traceability.

\paragraph{Known solution}\label{axiom:p} Another important instance of~\Cref{ruleform:e} is when $n=0$, i.e., when the change problem has a known solution:
\begin{example}
\[\begin{prooftree}
		\hypo{\qquad}
	\infer1[\using[Known solution]{$F$ satisfies $G$, implementation of $F$ satisfies $G$}]
		{E\update F\meets{G} N}
\end{prooftree}\]
meaning
\blockquote{$F$ is a solution to change problem $(E,N)$ if $G$ is satisfied with it and its implementation.}
\end{example}

\begin{exampleA}[Phone upgrade, cont'd]
	We can transform the original problem as follows into separate subproblems:
	\[\begin{prooftree}
	\hypo{\begin{array}{lll}
	[Phone] \update F \meets{G} New\_Phone\\
	\quad [Phone, Funds, Recycle]\update F \meets{G} New\_Phone\\
	\qquad [Phone, Tech, Physical]\update F \meets{G} New\_Phone\\
	\qquad\quad [Phone,Diary]\update F \meets{G} New\_Phone
	\end{array}}
\infer1{
	[Phone, Funds, Tech, Physical, Recycle,Diary]\update F \meets{G} New\_Phone}
	\end{prooftree}\]
The reader will note that:
\begin{itemize}
\item  $Funds$ and $Recycle$, and $Tech$ and $Physical$ are kept together as there may be recycling scheme which included trade in which would reduce the $Funds$ needed; it may also be that cables and cases can be transferred across, too
\item that $Phone$, $New\_Phone$, and $F$ (and $G$) are repeated in each subproblem, ensuring that the same solution is found in each subproblem (with each subproblem solution being validated by the same stakeholder): it is said that these problems \emph{tangle} together. Thus, the simple choice of $New\_Phone$ in the first subproblem must be a solution to each -- it must
	\begin{itemize}
	\item be affordable and recyclable, perhaps with a trade-in value;
	\item re-use cables and cases \wrt{} the stakeholder;
	\item be available in the diary window.
	\end{itemize}
\end{itemize}
\end{exampleA}

\subsection{Change problem rule classes}

We can identify specific classes of change problem rules according to their use within change problem solving. In particular, there are two classes of transformation: those that record improved understanding of a change problem, 
	and those for change engineering. 
	
\subsubsection{Problem understanding}\label{problemUnderstanding:s}

This section introduces the set of formal rules in \POED{} that support the progressive deepening of problem understanding as part of the change engineering process. As rule classes, we name the rule, the name being written above the justification/implementation part. 
The problem understanding rules facilitate modifications to existing problem expressions, such as the substitution or enhancement of environmental and need descriptions. This iterative approach to redefining the context and requirements of a problem allows for an evolving representation that reflects a more precise and nuanced understanding of, what is potentially, the complex and volatile change landscape. In employing these rules, the change engineer will systematically manage the evolution of problem descriptions, reducing ambiguities and aligning solutions with stakeholder perspectives and contextual shifts.

We note that the following problem understanding rules may appear to allow any change. To do so, however, would require the stakeholder to be entirely uncritical. While such stakeholders theoretically exist, they will not in any real-world organisational scenarios.

In the worst case, in situations of external change, a change in the organisational environment or need may require the reconsideration of all previous problem solving activity. Their application thus raises the risk of problem solving resource loss, which must be managed. The opportunity to do so is part of the macro-process below.

\paragraph{Environment Refinement}\label{envRef:p}

Environment refinement involves modifying the existing environment to incorporate necessary changes in the understanding of the Environment. 

\NewDocumentCommand\ruleenvexp{O{E} O{E'} O{F} O{N} O{G} O{$J$ satisfies $G$, $I$ satisfies $G$}}{
\begin{prooftree}
\hypo{#2  \update  #3 \meets{#5} #4}
\infer1[\using[Environment refinement]{#6}]
{#1  \update  #3 \meets{#5} #4}
\end{prooftree}}
\[\ruleenvexp\]
Examples of environment refinement include, for instance, the addition of a previously unrecognised domain or phenomena relationship. Changes to domains and phenomena outside of the organisation can also be captured by this rule.

\paragraph{Need Refinement}\label{needRef:p}

Need refinement patterns capture the evolution of the need description as the change problem is better understood. This involves refining and expanding the need to ensure it accurately reflects the desired outcome.

\NewDocumentCommand\ruleneedexp{O{N} O{N'} O{F} O{E} O{G} O{$J$ satisfies $G$, $I$ satisfies $G$}}{
\begin{prooftree}
\hypo{#4  \update  #3 \meets{#5} #2}
\infer1[\using[Need refinement]{#6}]
	{#4  \update  #3 \meets{#5} #1}
\end{prooftree}}
\[\ruleneedexp\]

Examples of need refinement include, for instance, the addition of a previously unrecognised requirement as a conjunction to the current need. Changes to needs arising outside of the organisation can also be captured by this rule. 

\paragraph{Change solution Refinement}\label{solnRef:p}

Change solution refinement captures the evolution of the solution as understanding develops:
\NewDocumentCommand\rulechangeexp{O{F} O{F'} O{E} O{F} O{G} O{$J$ satisfies $G$, $I$ satisfies $G$}}{
\begin{prooftree}
	\hypo{#3  \update  #2 \meets{#5} #4}
	\infer1[\using[Solution refinement]{#6}]
		{#3  \update  #1 \meets{#5} #4}
\end{prooftree}}
\[\rulechangeexp\]
Examples of change solution refinement include, for instance, the sequencing of a staged change solution (see~\Cref{macro:s}).

\subsubsection{Change engineering}\label{changeEng:ss}

The following rules correspond to~\cref{adddom:e,refinedom:e,cancel:e,changeable:e,need:e} of~\Cref{POE-Delta-Grammar:defn}.

\paragraph{Domain Addition}\label{CEDA:ss}

This transformation describes the case in which a step of exploration leads to the proposed addition of a domain

\newcommand\ruleChangeExpressionDomainAddition{
\begin{prooftree}
	\hypo{E(C)\meets{G} N}
\infer1[\using[Domain Addition]{$true$, Schedule $C$'s addition}]
	{E \update \domadd{C} \meets{G} N}
\end{prooftree}}
\[\ruleChangeExpressionDomainAddition\]

\newcommand\ruleChangeExpressionDomainAdditionExample{{\small
\begin{prooftree}
	\hypo{[sales, HR, finance, marketingTeam](marketingTeam)\meets{Manager} IncreasedRevenue}
\infer1[\using[Domain Addition]{$true$, Schedule $C$ addition}]
	{[sales, HR, finance]\update \domadd{marketingTeam})\meets{Manager} IncreasedRevenue}
\end{prooftree}}}
\begin{exampleA} Adding a marketing team:
\[\ruleChangeExpressionDomainAdditionExample\]
\end{exampleA}

\paragraph{Domain Removal}\label{CEDRem:ss}

This transformation describes the case in which a step of exploration leads to the proposed removal of a domain. The resulting problem is solved if the environment, $E$, without $D$ solves the problem.

\newcommand{\ruleChangeExpressionDomainRemoval}{
\begin{prooftree}
	\hypo{E(\nulldomain) \meets{G} N}
\infer1[\using[Domain removal]{$true$, Schedule $D$'s removal}]
	{E(D)  \update \cancel{D} \meets{G} N}
\end{prooftree}}
\[\ruleChangeExpressionDomainRemoval\]

\newcommand{\ruleChangeExpressionDomainRemovalExample}{
{\small\begin{prooftree}
	[sales, HR, finance](\nulldomain) \meets{Manager} OptimisedOperations
\using{\how [\textwidth]{Domain removal}[$true, \textrm{Remove $redundantDepartment$}$]}
\justifies
	[sales, redundantDepartment, HR, finance]\update\cancel{redundantDepartment} \meets{Manager} OptimisedOperations
\end{prooftree}}}
\begin{exampleA} Removing a redundant department:
	\[\ruleChangeExpressionDomainRemovalExample\]
\end{exampleA}

\paragraph{Domain refinement}\label{CEDref:p}

The complex change is the removal or refinement of a number of domains, $D^i$, implemented together, together with the addition of others, $C_j$. The general form of a complex change expression, in the context of a change problem, is
\begin{equation}
\begin{prooftree}
	\hypo{E(\astruct[D_1,...,D_n](C_1,...,C_m)) \meets{G} N}
\infer1[\using[Domain refinement]{$true$, Schedule $D$'s refinement}]
	{E(D) \update D\becomes \delta[D_1,...,D_n](C_1,...,C_m)\meets{G}N}\label{complexce:e}
\end{prooftree}
\end{equation}%
where $\astruct$ is the architecture introduction operator of \POE{}\textcite{hall2017a-design}.

\newcommand{\ruleChangeExpressionDomainRefinementExample}{
{\small\begin{prooftree}
{\begin{array}{ll}
	[sales, ITStructure, HR, finance](\astruct[restructuredSales](customerSupport))\\
	\qquad\meets{Manager} ImprovedEfficiency \\[1ex]
\end{array}}
\using{\how [\textwidth]{Complex change}[$true, \textrm{Removal, addition, refinement}$]}
\justifies
{\begin{array}{ll}
	[sales, ITStructure, HR, finance]\\
	\qquad\update sales \becomes \delta[restructuredSales](customerSupport) \\
	\qquad\qquad\meets{Manager} ImprovedEfficiency
\end{array}}
\end{prooftree}}}
\begin{exampleA} We refine the current $sales$ department to include a new $customerSupport$ function.
	\[\ruleChangeExpressionDomainRefinementExample\]
\end{exampleA}


\paragraph{Parallel change}\label{CECP:ss}

This transformation 
	describes the case in which change can be spilt into two disjoint contexts, i.e., in which there are no shared phenomena. 	
	Neither parallel not sequential change (to follow) transform directly to \POE{} problems as they have implementation implications (the $I$) which are not representable in \POE{}. 

\newcommand{\ruleChangeParallel}{
\begin{prooftree}
	\hypo{E_1 \update F_1 \meets{G} N_1} 
	\hypo{E_2 \update F_2 \meets{G} N_2}
\infer2[\using[Parallel]{$E_1\cap E_2=\emptyset$, Schedule $I_1$, $I_2$}]
	{(E_1\cup E_2) \update (F_1\parallel F_2) \meets{G} N_1\parallel N_2}
\end{prooftree}}
\[\ruleChangeParallel\]

\noindent The phenomenal independence of $E_1$ and $E_2$ should be validated by $G$.

\begin{exampleA}
\todo{Expand}solve(Manager, [HR, IT, finance], 'BetterPerformance \&\& IncreasedSecurity', 
      improveHR || upgradeIT, J, I) :-
    solve(Manager, [HR], 'BetterPerformance', improveHR, J1, I1),
    solve(Manager, [IT], 'IncreasedSecurity', upgradeIT, J2, I2),
    combineJustifications(J1, J2, J),
    combineImplementations(I1, I2, I).
\end{exampleA}

\paragraph{Sequential change}\label{CECS:p}

This transformation 
	describes the case in which change is sequenced through an intermediate change expression

\newcommand{\ruleChangeSequence}{
\begin{prooftree}
	\hypo{E \update F_1 \meets{G} N_1}
	\hypo{(E \update F_1)\update F_2 \meets{G} N_2}
\infer2[\using[Sequence]{$true$, Schedule $I_1$ then $I_2$}]
	{E \update (F_1\seq F_2) \meets{G} N_1\seq N_2}
\end{prooftree}}
\begin{equation}
\ruleChangeSequence\label{CECS:eq}	
\end{equation}

Sequencing requires the identification of an intermediate organisational need, $N_1$, which is satisfied by $E\update F_1$ and from which the final need, $N_2$, is satisfiable by the changes in $F_2$. The difficulty with sequence is that, should the final need not be satisfiable from an organisation in the intermediate state, then the whole intended change will be lost.

\paragraph*{Sequential domain refinement}\label{CESeqDref:p} 
A special case of sequencing is allowed under domain refinement with the following equivalence:
\begin{equation}
D\becomes\delta[D^1](C^1)\seq \delta[D^2](C^2)\equiv (D\becomes\delta[D^1](C^1))\becomes \delta[D^2](C^2)\label{CESeqDref:e}
\end{equation}
where $D^i$, $C^i$ are collections of domains.

\paragraph{Solution reflection}\label{CESolRefl:p}

To introduce parallel and sequence, environment, need, and solution refinement combine, for example, given some functional combination of parallel and sequence $f$ on needs $N_1,...,N_n$, we can reflect that structure into the solution space, thus:
\def\ruleneedexp{
\begin{prooftree}
\hypo{E  \update  f(F_1,...,F_n) \meets{G} f(N_1,...,N_n)}
\infer1[\using[Solution reflection]{$true$, Schedule the $F_i$ according to $f$}]
	{E \update  F  \meets{G} f(N_1,...,N_n)}
\end{prooftree}}
\[\ruleneedexp\label{solnRefl:e}\]

\begin{exampleA}
Example or two here (one of which should be low hanging fruit that breaks the transformation).

\todo{Expand}solve(CEO, Organisation, 'EnhancedProductivity', 
      reorgPhase1;reorgPhase2, J, I) :-
    solve(CEO, Organisation, 'InterimGoalsMet', reorgPhase1, J1, I1),
    applyChange(Organisation, reorgPhase1, OrgAfterPhase1),
    solve(CEO, OrgAfterPhase1, 'EnhancedProductivity', reorgPhase2, J2, I2),
    combineJustifications(J1, J2, J),
    combineImplementations(I1, I2, I).
\end{exampleA}

We claim that these eight rules formally characterise all possible change operations, i.e., that any required changes necessary to satisfy a need can be fully represented and effectively addressed using only combinations of them.

\subsubsection{Micro-process: summary} 


\POED{}'s micro-process is a significant contribution to the management and coordination of complex changes as it facilitate the detailed structuring of relationships, guides the nuanced decomposition of problems, manages interactions between actors at a granular level, and maintains coherence throughout each step of the problem-solving process. By providing this fine-grained support, \POED{} empowers change engineers working above it to effectively navigate the detail of organisational change, including meticulous risk management and ensuring that every component of the change process is thorough, transparent, and precisely aligned with the organisation's strategic objectives.

\subsection{The relationship between brown- and greenfield problem solving}

Three rules -- domain addition (\Cref{CEDA:ss}), domain removal (\Cref{CEDRem:ss}), and domain refinement (\Cref{CEDref:p}) -- have, as premises, \POE{} problems. These rules forge the link between the two frameworks, and determine the relationship between brown- and greenfield problem solving.

For domain addition, the associated \POE{} problem, once solved, implements its solution simply by installing it, but without otherwise altering, its environment. For domain removal, the associated \POE{} problem \enquote{installs} the empty domain, $\nulldomain$, which satisfies the need if and only if $E$ does. For domain refinement, the associated \POE{} problem uses an AStruct\textcite{hall2017a-design} by which domain refinement in the greenfield setting is achieved.

Because each changes the environment, each such rule application introduces a new \enquote{phase} of problem solving in which a different environment, $E_0, ..., E_n$, is to be changed. In the worst case, such environmental changes will rigorously enforce the linearisation of change, preventing later changes to be begun before earlier ones have completed. There are two mitigating factors:
\begin{itemize}
\item through analysis, the linearisation may not be strictly necessary, as when the environment can be partitioned into non-interacting parts; and
\item by appeal to the validating stakeholder, the risks of a non-strict linearisation might be borne by them, rather than the delegate.
\end{itemize}
However, there are also complicating factors such as, for example, when a forward dependency exists between changes, so that some information about a later change is needed to change an earlier part. This situation can be controlled by characterising the later change in some way, the phenomena it will affect, for instance, without knowing the details of precisely how they will be affected\footnote{Although not done in \POED{}, there is an example of such change in\textcite{hall2017a-design}.}.

All this is modulo stakeholder, of course, who will have the responsibility for the validation of the phenomenal analysis.

\subsection{The macro-process: delegated problem solving}\label{macro:s}

%

The change problem-solving process in \POED{} involves a series of large steps that explore, rationalise, validate, and design the change and the implementation of that change. This process is iterative and occurs whenever there is delegation in change engineering, as might be the case when a change problem owner $G$'s delegates $D_i$ to a change engineer which may occur many times for any change problem. The macro-process is designed to transfer any problem solving risks from the change engineer to the change problem owner. 


Delegation brings with it the following Change Problem Solving (CPS) steps for the delegate(s):
\begin{enumerate}[itemindent=1cm,label=\bfseries CPS\arabic*]
	\item \label[CPSstep]{CPS1}\textit{Change Problem Exploration}: $D_i$ creating their own view, $(\Sigma_{D_i}, N_{D_i})$, of $G$'s change problem\footnote{It may, of course, be that the $D_i$ work together to achieve a shared understanding, but this is not assumed.}

	\item \label[CPSstep]{CPS2}\textit{Change Problem Validation}: $D_i$ requesting validation from $G$ that $(\Sigma_{D_i}, N_{D_i})$ is an adequate interpretation of $(\Sigma_G,N_G)$. 
\end{enumerate}
	If validation given then~\Cref{CPS3}, otherwise~\Cref{CPS1}.
\begin{enumerate}[itemindent=1cm,label=\bfseries CPS\arabic*,resume]
	\item \label[CPSstep]{CPS3} \textit{Change Solution Exploration} $D_i$ identifying a new environment $\Sigma_{D_i}\update F_i$, which consists of 
	\begin{enumerate}[label=(\roman*)]
		\item identifying those the parts of $\Sigma_{D_i}$ that can remain unchanged, together with
		\item an implementation path for the solution $F_i$ with $\Sigma_{D_i}$, and 
	\end{enumerate}

    \item \label[CPSstep]{CPS4}\textit{Change Solution Validation}: $D_i$ requesting validation from $G$ that:
		\begin{itemize}
		\item that $J_i$ is adequate justification that $\Sigma_{G}\update F_i$ meets the agreed recognised need $N_{G}$;
		\item the implementation path $I_i$ is feasible.
		\end{itemize}
\end{enumerate}
	If validation given then~\Cref{CPS5}, otherwise~\Cref{CPS3}.
\begin{enumerate}[itemindent=1cm,label=\bfseries CPS\arabic*,resume]
 \item \label[CPSstep]{CPS5}\textit{Change Solution Implementation}: either:
\begin{enumerate}[label={\roman*)}]
	\item  migrate from $\Sigma_{D_i}$ to $\Sigma_{D_i}\update F_i$ using the validated implementation path, or 
	\item pass the implementation of $F$ back to the original problem owner to be combined into their implementation, potentially through a further implementation delegate.
\end{enumerate}
\end{enumerate}


\Cref{CPS1,CPS2,CPS3,CPS4} are similar to their \POE{} counterparts:
\begin{enumerate}[itemindent=1cm,label=CPS\arabic*:]
\item [\Cref{CPS1}]The initial \textit{Problem Exploration} step serves for the change delegate to develop their understanding of the change problem owner $G$'s scenario, \emph{viz}.~developing their version of $\currentstate$ and $N$, through discussions, review of available documentation, or any other appropriate means. 

\item [\Cref{CPS2}] given a candidate problem scenario $(\currentstate ,N)$, through validation, the delegate controls the risk that they are addressing the wrong scenario, at least \wrt{} the problem owner. This can be done as formally or as informally as the context requires. 

\item [\Cref{CPS3}] on validation of the delegate's change problem scenario, the delegate can begin their search for a solution, to identify $F$ in $\currentstate\update F\meets{G} N$. This step may involve further delegation of the problem or its parts to other delegates, whence the macro process will recommence for each, with the delegate in the role of problem owner.

\item [\Cref{CPS4}] given a candidate solution $F$, the delegate will seek validation that the problem owner sees the solution as fit-for-purpose for $\currentstate$.

\item [\Cref{CPS5}] This case is not part of the \POE{} macro-process and, here, the relationship between the \emph{problem owner}, the \emph{problem-solving delegate}, and the potential \emph{implementation delegate} becomes pivotal: the recursive nature of delegation means that while the problem owner entrusts the problem-solving delegate with devising a solution to meet a specific need, the actual implementation of this solution may be further delegated. In all but the most trivial of cases, therefore, the problem-solving delegate would not execute the implementation themselves.

Thus, when the problem-solving delegate is not be equipped or authorised to execute the implementation themselves, an \emph{implementation delegate} will be appointed to carry out the practical execution of the solution. The implementation delegate receives the validated solution and implementation plan from $D$ and is responsible for enacting the changes to transition the organisation from the current state $\currentstate$ to the updated state $\currentstate \update F$. The implementation delegate may further delegate specific tasks, adding additional layers to the delegation hierarchy, as described in~\Cref{delegate:p}.

The recursive delegation will continue until the responsibility for implementation reaches a party equipped to execute the changes effectively. This structure allows for specialisation, with each delegate focusing on their area of expertise-strategic planning by the problem owner, analytical problem-solving by the problem-solving delegate, and practical execution by the implementation delegate. 


By clearly defining these relationships and responsibilities, organisations can navigate the complexities of change implementation. This approach ensures that the solution is not only theoretically sound but also practically viable, facilitating a smoother transition and achieving the desired outcomes of the organisational change.
\end{enumerate}

Concurrent with the execution of steps \Cref{CPS1,CPS2,CPS3,CPS4,CPS5}, each delegate must be aware that the context of the change may change, so each step brings with it the requirement to monitor the environment and need, ideally in collaboration with the problem owner.

\subsection{Integrating micro- and macro-processes}

To extent necessary, determined by the level of abstraction (which, in turn, is determined by the criticality of the change (see~\Cref{changeexpr:s}), the micro- and macro-process must align with each other in their application. 
To do so, both justification ($J$) and implementation ($I$) components of rule applications must be embedded within application of the macro-process, as shown in~\Cref{JIReq:t}.

{\begin{tblr}[long,caption = {Requirements for justification $J$ and implementation $I$ across micro- and macro-processes},label = JIReq:t
]{
	colspec = X[-1]X[40]X[40],
	row{1} = {font = {\bfseries\normalsize}, c},
	row{even} = {font = \bfseries}
	}
Area&Justification ($J$) requirements&Implementation ($I$) requirements\\\hline
\SetCell[c=3]{l}Coordination of Delegation\\
&$J$ must include a clear rationale for how subproblems are divided and assigned, ensuring that each delegate understands the broader goals and constraints of the change. $J$ must capture the validation criteria for each delegate's solution to ensure alignment with the primary problem owner's objectives. The coordination element ensures that each subproblem's solution rationale is interconnected within the overarching $J$.& No requirements\\

\SetCell[c=3]{l}Integration of Results\\ 
	&$J$ justifies the compatibility and alignment of solutions from different subproblems. The combined $J$ should outline how the sub-solutions relate and contribute to the overall organisational problem. 
	&The implementation plan ($I$) must specify how these results will be integrated into a coherent whole, detailing the sequence and conditions required for combining these solutions effectively.\\

\SetCell[c=3]{l}Management of Dependencies\\ 
&$J$ should justify the sequencing itself, arguing how dependencies require the specified implementation paths.
	&$I$ must clearly document dependencies between subproblems and the conditions under which they can be implemented. This may include constraints on the order of implementation (e.g., the  implementation of $F_1$ must precede that of $F_2$ if so required).\\

\SetCell[c=3]{l}Risk Management \\
&$J$ should capture any risks associated with the delegated subproblems and provide justification for their mitigation. $J$ should also outline where in the validation component the mitigation of risks will have been performed.
	&$I$ should detail the risk management steps within the implementation plan, including contingency actions and checkpoints that allow for proactive responses to issues that may arise during implementation.\\

\SetCell[c=3]{l}Feedback Loops and Iterative Adjustments\\ 
	&$J$ should recognise the backtracking involved in maintaining alignment with the problem owner's needs as the problem-solving and implementation phases progress in potentially volatile contexts. The tempi at which this should be done should match the volatility of the context\textcite{costantini2021using}. 
		& $I$ should incorporate steps for gathering feedback, revisiting sub-solutions where dependencies necessitate them, and making iterative changes based on new insights. This recursive nature ensures that the macro-process can adapt dynamically to evolving conditions or stakeholder input.\\

\SetCell[c=3]{l}Alignment of Implementation Timelines\\ 
& $J$ can provide the justification for why specific timelines were chosen, ensuring that they align with the overall strategic goals and practical limitations.
	& $I$ must outline the scheduling of each sub-solution, considering dependencies and available resources. $I$ should include sequencing constraints (e.g., sequential or parallel application rules) to maintain coherence across subproblems.\\

\SetCell[c=3]{l}Validation and Assurance\\ 
	&$J$ should document the criteria for validating each sub-solution and how these validations align with the overarching organisational need. Each delegate's $J$ must include detailed arguments supporting the sub-solution's fitness for purpose. 
	& $I$ should outline the process for performing these validations during and after implementation, ensuring that each subproblem contributes effectively to the final outcome.\\

\SetCell[c=3]{l}Resource Allocation and Management \\
	 &$J$ should justify the allocation of these resources, demonstrating their alignment with the priorities and constraints of the overall change initiative.&No requirements
\end{tblr}}

\subsection{Bounding change and its analysis}

In organisational and systems change management, modification complexity is increased by the existence of dependencies, the need to maintain continuity during transition to the new system, and the risk of unintended consequences on overall performance. Addressing these challenges requires a structured approach that decomposes changes into manageable, logically ordered steps. Traditional approaches provide various methods for  structured change \parencite[see, for instance]{vincenti1990what}. 

In the redesign of any complex organisation to meet organisational needs, therefore, it is crucial to manage and limit the scope of changes to minimise organisational disruptions. By modelling the organisation as a set of domains defined by phenomena and their causal interactions, we can analyse how changes propagate and identify ways to bound these changes within specific domains.

When we design a domain, we alter its description -- these are called structural changes. Another form of change is  behavioural change, which are changes in the occurrences of phenomena without altering the domain's structure.

During change analysis for any particular domain, we must aim to be as parsimonious with change as is concomitant with 
%
the need to change;
the need to restrict change to those areas of the organisation that are permitted to change
%
and so we must consider both direct and transitive effects between domains. 

A transitive effect occurs when a change in one domain affects another domain indirectly through intermediate domains, leading to cascading effects throughout the organisation. For example, suppose domain $C$ is changed to introduce a new causal relationship between phenomena $y$ and $c$, so that $y \rightarrow c$ which did not exist before\footnote{Using $\rightarrow$ to express causality\textcite{moffett1996model}.}. This change may not structurally alter a domain $D$ that observes either $y$ or $c$, but it can cause new occurrences of phenomenon $c$ in $D$ (those now caused by $y$), leading to behavioural changes therein. If domain $D$ has a causal relationship $c \rightarrow d$, and domain $E$ has $d \rightarrow f$, the initial change in domain $C$ can propagate through each of these, resulting in unintended -- or emergent -- behavioural changes elsewhere.

To effectively bound changes, we therefore associate with each domain, $D$, two sets of phenomena: those observed ($D_o$) and controlled ($D_c$) by $D$. These are available, on inspection, to the change engineer, even though the (internal) description of the domain may not be. We analyse the observed and controlled phenomena in each domain to determine how changes may propagate. This involves:
\begin{enumerate}
\item Identifying the phenomena that each domain observes and controls.
\item Mapping the causal relationships between phenomena within and between domains.
\item Assessing whether intermediate domains act as \emph{change buffers}, absorbing changes without propagating them further.
\end{enumerate}
In the case that a change buffer exists for a particular phenomenon, we complete our change analysis for that phenomenon. In the case when transitive change is felt, other options can be considered, such as remaining phenomena for that there is no capture by other domains' observed or controlled phenomena. In the worst case, however, another step in a transitive analysis should be completed.

By understanding these relationships, the change engineer can make informed decisions to isolate structural changes, coordinate between domains when necessary, and anticipate cascading effects. This careful analysis helps to limit the scope of changes, maintain organisational stability, and ensure that the redesign meets the new needs effectively without unintended consequences.

Bounding change within an organisation involves a detailed examination of domain interactions through phenomena and causal relationships. By anticipating transitive effects and potential cascading impacts, we can strategically plan changes to confine them to specific domains or manage them across domains with minimal disruption. This approach enables organisations to adapt to new requirements while preserving the integrity and performance of existing systems.
 
\subsection{Extended illustrative example}

We explore the relation between micro- and macro-processes through an example. 

Our example focuses on upgrading a development environment's API, where existing libraries, tools, or other dependencies must be managed carefully to ensure that updates do not compromise functionality in dependent applications. Given that there exists an installed based of $OldAPI$-based applications, $G$, the problem owner, wants to avoid a single step update in the API, implementing a 2-stage approach\parencite{sturgeon2020deprecating}. 

This situation demands an incremental, two-stage upgrade approach that balances stability and enhancement. The strategy involves initially deprecating the older API while keeping it active to maintain continuity, allowing the new API to run in parallel. Once compatibility is ensured and system requirements are met, the older API can safely be removed. This phased approach minimises risks and provides testing and verification opportunities at each stage, aligning with organisational goals for reliability and advancement.

The problem is formalised in \POED{} as the need, $UpdateAPI$, within the organisational development environment, $DevEnv$, which includes domain $OldAPI$. We will assume that the problem owner is $G$. The change scenario is:
\[(DevEnv(OldAPI),UpdateAPI)\]
with associated change problem:
\begin{equation}
	DevEnv(OldAPI)\update F\meets{G} UpdateAPI \label{orig}
\end{equation}
which, according to~\Cref{POED:eq}, means:
\blockquote{Starting from the existing environment $DevEnv(OldAPI)$, find implementable changes $F$ such that the updated environment $DevEnv(OldAPI)\update F$ meets the need $UpdateAPI$ \wrt{} problem owner $G$}

The delegate rule (\Cref{delegate:p}) allows us to delegate to $G$'s trusted delegate, 
\[
DevEnv(OldAPI)\update F\meets{D} UpdateAPI
\]

$D$ then uses need refinement (\Cref{needRef:p}) to identify an intermediate point, $UpdateAPI^{addNew}_{deprecateOld}$, in the replacement: 
%
\begin{equation}
	DevEnv(OldAPI)\update F\meets{D}UpdateAPI^{addNew}_{deprecateOld};UpdateAPI\label{2stageneed}
\end{equation}
%
%
which we can reflect into the solution space using solution reflection (\Cref{CESolRefl:p}):
\begin{equation}
	DevEnv(OldAPI)\update F_1\seq F_2\meets{D}UpdateAPI^{addNew}_{deprecateOld};UpdateAPI\label{need2soln}
\end{equation}
%
%
identifying two subproblems, one for each stage of the change, that are related by their shared references to domains $DevEnv$, $OldAPI$, and $F_1$:
\begin{equation}
	\textrm{Stage 1:}\qquad DevEnv(OldAPI)\update F_1\meets{D}UpdateAPI^{addNew}_{deprecateOld}\label{stage1}
\end{equation}
and
\begin{equation}
	\textrm{Stage 2:}\qquad(DevEnv(OldAPI)\update F_1)\update F_2\meets{D}UpdateAPI\label{stage2}
\end{equation}
via sequence refinement (\Cref{CECS:p}). 

For Stage 1, we will:
\begin{itemize}
\item  replace $OldAPI$ with a version $OldAPI'$ that prints a deprecation message on use during compilation, and 
\item introduce the $NewAPI$
\end{itemize}
captured via domain refinement (\Cref{solnRef:p}) of $OldAPI$ thus:
\begin{equation}
	DevEnv(OldAPI)\update OldAPI\becomes\delta[OldAPI'](\domadd{NewAPI})\meets{D}UpdateAPI^{addNew}_{deprecateOld}
\end{equation}
which is a greenfield and, hence, the \POE{} problem (\Cref{CEDref:p}):
\begin{equation}
	DevEnv(\astruct[OldAPI'](NewAPI))\meets{D}UpdateAPI^{addNew}_{deprecateOld}\label{stage1.1}
\end{equation}
%

For Stage 2, we substitute back for $F_1$ into~\Cref{stage2}, and refining $F_2$ with $\cancel{OldAPI'}$ gives: 
\begin{equation}
	DevEnv(OldAPI',NewAPI)\update \cancel{OldAPI'}\meets{D}UpdateAPI\label{stage2.1}
\end{equation}
which is another greenfield problem (\Cref{CEDref:p}):
\begin{equation}
	DevEnv(NewAPI)\meets{D}UpdateAPI\label{stage2.2}
\end{equation}
in which we are left with the \POE{} problem to design $NewAPI$.

Putting everything together, we have the whole solution:
\begin{equation}
	DevEnv(OldAPI)\update OldAPI\becomes\delta[OldAPI'](\domadd{NewAPI})\seq \cancel{OldAPI'}\meets{D}UpdateAPI\label{recompose}
\end{equation}

In this example, the timings are currently implicit. In reality, they would be sequenced through consecutive API changes, placing timing and other constraints on the stages and when they must be completed. This information is contained in the implementation element of the framework.

There are a number of notable characteristics of this change:
\begin{itemize}
\item the sequence refinement step constrains the implementation $I$ such that $F_1$ is to be installed before $F_2$. As mentioned in~\Cref{CPS5}, we should be clear what this implies for our problem solving process. We note that it does not mean that any work on $F_2$ should be prevented before that on $F_1$ is complete -- this would be too strict a condition and might lead to $F_1$ being unsolvable when it depends on a particular detail of $F_2$ -- a communication mechanism, for instance. In fact, in the worst case, $F_1$ will be undeliverable without all work on $F_2$ being completed. The constraint is, therefore, only on the order in which their implementation is delivered.
	 
\item the first step is to refine the need ($UpdateAPI$ becomes $UpdateAPI^{addNew}_{deprecateOld};UpdateAPI$) to introduce the staging of the update which is then reflected in the solution ($F$ becomes $F_1\seq F_2$). Indeed, we conjecture that complex refinements of a need in terms of sequential and parallel composition can be reflected in the solution, to identify a solution component associated with each sub-need.

\item although we do not do it here, staging would typically be timed. This, we could associate with the sequencing an absolute date (\texttt{2024-11-01}, for instance) or relative (for inclusion with update \texttt{10.3.2}, for instance) at which this must take place.

\item even on this simple example, we have gone into a great deal of detail to illustrate the working of the micro-process. Applications of the macro process consistent with the micro-process would be sufficient in most simple cases.

\item there are many rule applications at each step, we have preferred a problem focus, working in the problem domain whenever possible.
\end{itemize}
In reality, given the complexity of a realistic development environment, it is highly unlikely that changes to the API would be isolatable within the API itself as many other components will need to be kept consistent, including core infrastructure, development and build tools, security and access control, logging and monitoring systems, user and developer permissions and there may be other dependencies unrelated to the API Layer. These will be isolatable due to their phenomenological independence from the API updates. On subsystems, those involving documentation will share phenomena and so need some level of updating as part of the API upgrade. In this case, it is the macro-process that provides the mechanisms by which this interdependency would be identified, either early in:
\begin{description}
\item [change problem exploration~\Cref{CPS1}] if the change delegate explored phenomenal relationships between code and documentation; 
\item [problem validation~\Cref{CPS2}] if the change problem owner required the change delegate to acknowledge the interdependence;
\end{description}
or later, in:
\begin{description}
\item [change solution exploration~\Cref{CPS3}] if the change delegate identified phenomenal relationships between code and documentation while exploring the solution; 
\item [change solution validation~\Cref{CPS4}] if the change problem owner was unable to validate a solution that did not recognise the interdependence;
\item [change solution implementation~\Cref{CPS5}] if the interdependence was revealed only during implementation.
\end{description}

With knowledge of this dependence, the \POED{} problem becomes:
\begin{equation}
	DevEnv(OldDoc, OldAPI)\update F\meets{D} UpdateAPI \label{origWithDoc}
\end{equation}
with solution
\begin{equation}
\begin{array}{l}
	DevEnv(OldDoc,OldAPI)\update OldAPI\becomes\delta[OldAPI'](NewAPI)\seq \cancel{OldAPI'}\\
	\qquad\parallel OldDoc\becomes\delta[OldDoc'](NewDoc)\seq \cancel{OldDoc'}\meets{D}UpdateAPI\label{recomposeWithDoc}
\end{array}
\end{equation}
which adds to the change solution the burden of ensuring the Documentation is up to date. 

In general, the earlier in the process that the issue is noted the lesser the risk, emphasising the importance of early problem exploration and validation as the basis of change management.

Of course, in the worst case, with a neophyte problem owner, delegate, or implementation team, the interdependence could have been missed throughout, to be revealed only on release to developers using the API; a costly mistake in terms of reputation.

\section{Evaluation}\label{sect:evaluation}

In Section~\ref{background:sec} we established some desirable characteristics of an holistic design-based approach to organisational change, to which we return in this section to provide our overall evaluation of \POED{} based on evidence from its application in our studies (see Table~\ref{studies:t}).

\paragraph*{Problem exploration} \POED{} promotes the exploration of the change need in context to motivate, inform, and instigate the change design process, and to provide the validation criteria for the designed change solution. This intrinsic rebalancing of the framework towards problem exploration is seen as a point of strength over approaches which favour moving quickly to solution, often prematurely. That this is a desirable characteristic of a design  problem solving framework was already established for its parent framework, \POE{}\parencite{hall2017a-design}, and is therefore maintained in \POED{}. We should note, however, that while a rebalancing towards the problem space is built into the framework, it does not prevent solution space exploration at any time. In fact, solution space exploration may even be desirable in situations where the change need is fuzzy to start with: by looking into the solution space we can achieve more clarity as to the change need in context the designed change should serve, as long as we avoid anchoring on a premature solution design which may prove unsuitable or even unfeasible in the long run. 

Evidence from the NICE case study demonstrates this: early discussions with stakeholders focused on capturing a shared understanding of the organisation's needs for their online presence, which proved instrumental in framing the problem correctly and helped discover hidden requirements and constraints. Similarly, in the OSLC study, stakeholders initially struggled to articulate integration requirements. By first exploring potential technical solutions, the team clarified the need to address cross-organisational dependencies and tool interoperability. This iterative interplay between need exploration and solution exploration highlights \POED{}'s effectiveness in managing complex, fuzzy change scenarios without anchoring to under-explored problems and, therefore, the potentially sub-optimal solutions arising.

\paragraph*{Identification of scope and boundaries}	
The boundaries of systems and sub-systems pertaining to the change problem can be established precisely in \POED{} through the definition of domains and their phenomena, with their key properties expressed as relationships between those phenomena, which also capture any interface and interdependence. This feature of \POED{} was successfully tested repeatedly in our studies which looked at change problems of different natures and increasing complexity. In particular, the OSLC case study presented a highly complex environment requiring the integration of various engineering tools across multiple organisations, each with distinct requirements, constraints, and technical sub-systems, so that the need to represent clearly and accurately those complex relationships was paramount. 


In the Air-Traffic Control (ATC) case study, the forensic analysis of past changes revealed overlooked dependencies that had contributed to previous project failures. By systematically re-establishing boundaries and interfaces, the retrospective analysis provided actionable insights to guide the subsequent design of a highly automated train dispatch platform. This demonstrated the utility of \POED{} in identifying scope and managing dependencies in both forward-looking and retrospective contexts. While the framework's capabilities to manage boundary complexities were confirmed, participants in the OSLC study also emphasised the cognitive demand of manually tracking these descriptions in large-scale projects, further underscoring the need for enhanced tool support.

\paragraph*{Capturing constraints} Related to scope and boundaries, \POED{}'s phenomenological basis allows the capture and analysis of constraints emerging from both context and design process, informing what is feasible in the solution space. In complex environments, like the OSLC study, those phenomenological relationships proved very effective in identifying deeply intertwined domains which prevented the application of divide-and-conquer solution strategies in favour of solution co-design -- situations known as \enquote{problem tangles} in \POE{} and \POED{}. In the presence of tangles, sub-problem decomposition needs careful management as changes cannot be isolated and implemented independently in different parts of the environment. 

The LLM-A case study highlighted this further, where the integration of a new AI-driven troubleshooting system required balancing constraints from legacy systems, user workflows, and regulatory requirements. The use of \POED{} facilitated the alignment of these constraints, ensuring that changes in one domain did not inadvertently disrupt others. Stakeholders specifically praised the framework's ability to maintain coherence in the face of overlapping and evolving constraints.

\paragraph*{Linguistic diversity} Different description languages can be used within the framework, with varying levels of formality and precision to suit different stakeholder needs and level of criticality of the change problem. This flexibility affords both the precision required in technical and highly critical contexts, and ease of communication and sense-making among non-technical stakeholders and problem owners. Feedback from study participants confirmed that being able to use different languages with different stakeholders was a highly desirable feature of the framework, and the studies we have conducted show how the framework is able to cope with such linguistic diversity. For example, the AI-A (Kettle) study both fully and less formal reasoning about safety-critical changes was employed, where precise and structured descriptions were essential to validating the impact of proposed changes on safety properties. Conversely, in the NICE study, informal sketches and textual descriptions were used effectively to communicate the scope of changes to non-technical stakeholders, emphasising accessibility and engagement.

\paragraph*{Reasoning}
Different description languages are brought together within the unifying reasoning capabilities of \POED{} through its phenomenological basis, so that phenomena relationships can be used as the basic mechanisms to reason about constraints and change impact across systems and sub-systems, even when these are described using different languages. This was demonstrated in particular in the AI-A study, where phenomena-based formal descriptions in and simulations derived from HCSP\textcite{chaochen1993extended,lv2013hcsp} were used to reason about the preservation of safety properties in the change from a manual to an AI-based controller. Note that the ability to reason about change impact through phenomena is essential in change problem solving: while in the greenfield world of \POE{} we could restrict all considerations of impact to the interface between an unchanged environment and the newly design artefact, in the brownfield world of \POED{} environmental change must be factored in at each step of the problem solving process, as potentially far-reaching environmental changes are concomitant to solving the change problem. 

In the OSLC study, reasoning based on phenomena relationships proved critical in managing dependencies between tools from multiple vendors, allowing the team to identify and address potential integration conflicts early. Similarly, in the LLM-A case study, reasoning supported the alignment of technical and organisational changes, ensuring that the new AI-driven system met operational requirements while minimising disruption to existing workflows.

\paragraph*{Diversity of perspectives}
Problems are expressed in relation to explicit validating stakeholders, with no expectation of a unified view by all, hence allowing different perspectives to be accounted for in the problem solving process. Of particular note is the distinction between problem owner and change designer formalised in the framework, which underlines the need to establish independent validation criteria for the designed solution, acknowledging the need to reconcile potentially divergent views of problem owner and solution designer. Such validation criteria are refined and factored into each step of the problem solving process through the justification and implementation conditions accompanying each problem transformation in the micro process, or their expression in the corresponding macro process.  

This flexibility was evident in the OSLC case study, where stakeholders across multiple organisations, each with their own priorities and constraints, participated in the design process. By explicitly capturing their diverse perspectives, \POED{} facilitated the reconciliation of these views to develop a shared understanding of the integration challenges. Similarly, in the NICE study, the framework enabled the accommodation of varied stakeholder inputs, from technical requirements to cultural and community-driven considerations, ensuring that the redesigned system met the collective goals of the group. Feedback from these studies highlighted the importance of structured validation steps in managing and harmonising divergent viewpoints.

\paragraph*{Design alternatives} Both micro and macro problem solving processes in \POED{} allow the exploration of alternative design paths, whose steps can be formally expressed, analysed and justified in relation to diverse stakeholders' validation criteria. This was particularly evident in the LLM-A case study, where iterative exploration of alternative AI system designs enabled stakeholders to identify the most viable approach. During initial trials, multiple configurations of the Retrieval-Augmented Generation (RAG) pipeline were tested to balance accuracy and system responsiveness, demonstrating the utility of a structured design exploration. Similarly, the ATC study underscored the importance of exploring design alternatives retrospectively, enabling the identification of suboptimal decisions in prior implementations that could inform improved practices in future projects. The ability to explore design alternative is of particular significance in organisational change: while in greenfield design, backtracking through design alternatives has little impact on the context, in brownfield design this may not the case as the sequentialisation of implementation steps may lead to irreversible environmental changes. 

\paragraph*{Assurance}	
Assurance is embedded in the problem solving process through validation criteria and step justifications, something that \POED{} has inherited from its parent framework, \POE{}\textcite{hall2009assurance-driven}, particularly in the context of safety critical systems and their statutory safety certification\parencite{mannering2008problem}.  \POED{} extends assurance arguments to the consideration of how implementation steps are sequenced, and may lead to intermediate environmental changes. Across all studies, assurance through recurrent validation steps emerged not only as a risk management tool but also as a mechanism to build stakeholder trust, enabling transparent decision-making and fostering confidence in the change process. Feedback from all case studies emphasised the utility of embedding assurance into each problem-solving step to manage risks effectively and provide traceable justifications for key decisions.

\paragraph*{Process support and guidance}	
Micro- and macro-processes in \POED{} are defined to support both formal and practical implementation of organisational change, including maintaining traceability of intermediate change design and implementation steps from initial change problem capture through its evolution during the problem solving process. They provide a systematic approach to change which is particularly effective in dealing with complex, multi-faceted change problems, as demonstrated in our studies, particularly the more complex OSLC and ATC studies. 

In the OSLC case study, the structured processes enabled stakeholders to trace interdependencies across multiple engineering tools and systems, facilitating more informed decision-making and risk management. To this end, the study participants, particularly expert engineers, indicated that they particularly valued the framework's structured processes and support for systematic problem solving and change impact analysis. 

In the ATC study, the framework was used to trace historic changes spotlighting past errors and providing actionable insights for future projects.  This demonstrated \POED{}'s utility as a forensic tool enabling the capture and analysis of past change processes, both to identify potential missteps which may have led to past failure, and effective change strategies for reuse in different change contexts, extending to change \POE{}'s idea of \emph{dactics}\parencite{hall2012software}, that is capturing design tactics for reuse. 

\paragraph*{End-to-end change process}	
From a theoretical standpoint, the integration between greenfield and brownfield design and implementation is enabled by \POED{} adopting the underlying \POE{}'s philosophical and formal foundation. Validating its effectiveness in practice was a major objective of the LLM-A study, which concerned the end-to-end transition from a manual system to an automated LLM-driven troubleshooting system in the context of a building automation company specialising in the installation and management of automation equipment in large industrial buildings. This included both technical changes, and organisational and process adjustments required to support the new system. The focus of the study was implementing the change and evaluating \POED{}'s ability to predict any wider effects within the organisation. Specifically, stakeholders noted that \POED{} was \enquote{very effective} in helping them identify \enquote{blind spots} concerning the integration of the new AI system within its operational environment. In particular, \POED{} helped the technical team predict required changes to existing workflows as, for example, the need for explainable AI and bespoke training for technicians to interpret and act on AI-generated insights, something the team had not considered in scope at the onset. 

Additionally, the LLM-A case study, which was our first to look at the entire span of change engineering, spanning not only the identification of the change domains but also the design of the new artefacts required to implement those changes, highlighted the iterative interplay between the \POED{} and \POE{} frameworks during the transition between change-centric and design-centric processes. One of the key observations was the need for re-alignment in granularity when transitioning from \POED{} to \POE{}. While \POED{} successfully identified the need for new artefacts (e.g., an AI-driven troubleshooting system), the subsequent \POE{} design process revealed the necessity of re-evaluating the environment, incorporating additional stakeholders, domains, and technical constraints that were not initially within the scope of the \POED{} analysis. For example, the greenfield design of the AI/LLM domain required the involvement of new roles, such as Data Scientists, Ontology Specialists, AI Architects, and Security Specialists, who brought with them emergent requirements that were not considered during the initial brownfield problem-solving phase. These findings suggest the importance of maintaining iterative loops between \POED{} and \POE{} processes, enabling dynamic refinement of both designs and change analyses under evolving constraints. As new stakeholders and domains emerge during the \POE{} design phase, their impact must be propagated back into the \POED{} analysis, ensuring that the broader organisational context remains aligned with the specific greenfield design. This iterative alignment shows the close link between the two frameworks.


\section{Conclusions and Future Work}

This article has introduced \POED{}, a novel change framework that offers support for the systematic capture, analysis, and solution of change problems in complex socio-technical contexts. \POED{} provides an end-to-end change problem-solving approach that integrates greenfield and brownfield development. Case studies validate it as a valuable tool for practitioners dealing with complex engineering change.

 \POED{} is novel in addressing gaps acknowledged in the literature in both high-level change management frameworks, which often lack practical implementation details, and purely design-focused approaches, which do not extend effectively into organisational change. Its structured processes and analytical and synthetic tools are validated through a number of empirical studies which have demonstrated utility and effectiveness of use within professional practice, particularly in their support for systematic and traceable change, stakeholder delegation and validation of change, and linguistic diversity within a unifying reasoning system.

While demonstrating \POED{}'s effectiveness and utility in supporting complex change engineering, our empirical studies have also highlighted areas that would benefit from further work, particularly in relation to enhancing \POED{}'s practicability for end-users, in particular, its usability and accessibility, viz.

Firstly, \POED{} would benefit from dedicated tool support, something highlighted by several stakeholders during the case studies. This would enable the framework's adoption and application in large-scale projects, where the use of manual tools to capture, develop and keep track of change problems is too burdensome. A useful tool should, at least, be able to capture and manage the framework's micro- and macro-process progress for problem-solving, particularly in environments where rapid changes and iterative validations are essential. This could be achieved with a suitably designed database backend for change problem solving, together with interrogative support for the change problem database.

Secondly, the integration of \POED{} with complementary methodologies and frameworks also warrants attention. In our studies, practitioners appreciated the ability to use \POED{} in combined with their current change processes, so that identifying effective ways to position the framework within industry-standard approaches, particularly in relation to managing complex change projects\parencite{costantini2021using}, could further extend its reach and adaptability. 

Lastly, further empirical validation is needed, through an expanded set of case studies across diverse sectors: this would further strengthen confidence in the framework as offering valuable assistance to the change engineer. While the case studies we have conducted have already shown \POED{} to be effective in various contexts, applying it within additional domains, particularly those with differing regulatory, technical, or cultural complexities, would enrich the evidence base, and provide the opportunity to validate how the framework performs when used by teams with varying levels of expertise in change management.

\raggedright
\printbibliography

\end{document}

%% file: localDefs.tex
\renewcommand{\seq}[0]{\mathbin{;}}

%% file: Bibliography/Refs.bib
@article{weick1999organizational,
	author = {Weick, Karl E and Quinn, Robert E},
	journal = {Annual review of psychology},
	number = {1},
	pages = {361--386},
	publisher = {Annual Reviews},
	title = {Organizational change and development},
	volume = {50},
	year = {1999}}

@techreport{lawler2015organization,
	address = {Marshall School of Business, University of Southern California Los Angeles CA 90089--0871},
	author = {Lawler, Edward E and Worley, Christopher G},
	institution = {Center for Effective Organizations},
	month = {4},
	number = {G15-10 (657)},
	pages = {1--21},
	title = {Organization agility and talent management},
	year = {2015}}

@inproceedings{tsiavos2021technology,
	author = {Tsiavos, Vasilis and Kitsios, Fotis},
	booktitle = {European, Mediterranean, and Middle Eastern Conference on Information Systems},
	organization = {Springer},
	pages = {681--693},
	title = {Technology as driver, enabler and barrier of digital transformation: A review},
	year = {2021}}

@incollection{rezgui2005socio-organisational,
	author = {Rezgui, Y., Wilson I. Olphert and Damodaran, L.},
	booktitle = {Virtual Organisations: Systems and Practices},
	publisher = {Springer Science+Busineess Media Inc.},
	title = {Socio-Organisational Issues},
	year = {2005}}

@book{chaochen1993extended,
	author = {Chaochen, Zhou and Ravn, Anders P and Hansen, Michael R},
	publisher = {Springer},
	title = {An extended duration calculus for hybrid real-time systems},
	year = {1993}}

@incollection{lv2013hcsp,
	author = {Lv, J and Li, K and Tang, T and Chen, L},
	booktitle = {Computers in Railways XIII},
	pages = {15--25},
	publisher = {WIT Transactions on The Built Environment},
	title = {{HCSP formal modeling and verification method and its application in the hybrid characteristics of a high speed train control system}},
	year = {2013}}

@inproceedings{mannering2008problem,
	address = {Bristol, UK},
	author = {Mannering, Derek and Hall, Jon G and Rapanotti, Lucia},
	booktitle = {Proceedings of the Safety-critical Systems Symposium 2008},
	month = 2,
	oro = {1},
	title = {Problem oriented safety process improvement},
	year = {2008}}

@article{hall2009assurance-driven,
	author = {Hall, Jon G. and Rapanotti, Lucia},
	journal = {International Journal On Advances in Systems and Measurements},
	number = {1},
	oro = {1},
	pages = {119--130},
	title = {{Assurance-driven design in Problem Oriented Engineering}},
	volume = {2},
	year = {2009}}

@inproceedings{hall2017a-phenomenal,
	author = {Hall, Jon G and Rapanotti, Lucia and Markov, Georgi},
	booktitle = {Proceedings of 5th IEEE International Workshop on Formal Methods Integration (IEEE FMi 2017)},
	oro = {1},
	title = {A phenomenal basis for hybrid modelling},
	year = {2017}}

@misc{sturgeon2020deprecating,
	author = {Sturgeon, Phil},
	howpublished = {https://blog.stoplight.io/deprecating-api-endpoints},
	lastchecked = {2024-10-31},
	month = {12},
	title = {Deprecating API Endpoints},
	year = {2020}}

@phdthesis{markov2024poe-delta,
	author = {Markov, Georgi},
	school = {School of Computing \& Communications, The Open University},
	title = {POE-$\Delta$: a framework for change engineering},
	year = {2024}}

@article{lewin1947frontiers,
	author = {Lewin, Kurt},
	doi = {10.1177/001872674700100103},
	eprint = {https://doi.org/10.1177/001872674700100103},
	journal = {Human Relations},
	number = 1,
	pages = {5--41},
	title = {Frontiers in Group Dynamics: Concept, Method and Reality in Social Science; Social Equilibria and Social Change},
	url = {https://doi.org/10.1177/001872674700100103},
	volume = 1,
	year = 1947}

@article{drucker1955management,
	author = {Drucker, Peter F},
	journal = {Management Science},
	number = {2},
	pages = {115--126},
	publisher = {INFORMS},
	title = {Management science and the manager},
	volume = {1},
	year = {1955}}

@book{kleene1964introduction,
	author = {Kleene, Stephen Cole},
	booktitle = {Introduction to Metamathematics},
	publisher = {North-Holland Publishing Company},
	title = {Introduction to Metamathematics},
	volume = 1,
	year = 1964}

@book{chin1969general,
	author = {Chin, Robert and Benne, Kenneth Dean and others},
	publisher = {Human Relations Center, Boston University Boston, MA},
	title = {General strategies for effecting changes in human systems},
	year = 1969}

@book{havelock1973change,
	author = {Havelock, Ronald G},
	publisher = {Educational Technology},
	title = {The change agent's guide to innovation in education},
	year = 1973}

@article{galbraith1974organization,
	author = {Galbraith, Jay R},
	journal = {Interfaces},
	number = 3,
	pages = {28--36},
	publisher = {INFORMS},
	title = {Organization design: An information processing view},
	volume = 4,
	year = 1974}

@article{zand1975theory,
	author = {Zand, Dale and Sorensen, Richard},
	journal = {Administrative Science Quarterly},
	number = 4,
	pages = {532--545},
	title = {{Theory of Change and the Effective Use of Management Science}},
	volume = 20,
	year = 1975}

@incollection{simon1977structure,
	author = {Simon, Herbert A},
	booktitle = {Models of discovery},
	pages = {304--325},
	publisher = {Springer},
	title = {The structure of ill-structured problems},
	year = {1977}}

@book{rogers1983nature,
	author = {Rogers, G. F. C.},
	publisher = {Palgrave Macmillan},
	title = {The Nature of Engineering: A Philosophy of Technology},
	year = {1983}}

@article{goel1989motivating,
	author = {Goel, Vinod and Pirolli, Peter},
	journal = {AI magazine},
	number = {1},
	pages = {19},
	publisher = {John Wiley \& Sons},
	title = {Motivating the notion of generic design within information-processing theory: The design problem space},
	volume = {10},
	year = {1989}}

@book{vincenti1990what,
	author = {Vincenti, Walter G.},
	publisher = {The Johns Hopkins University Press},
	title = {What Engineers Know and how they know it: Analytical studies from Aeronautical History},
	year = {1990}}

@incollection{funke1991solving,
	author = {Funke, Joachim},
	booktitle = {Complex problem solving},
	pages = {185--222},
	publisher = {Psychology Press},
	title = {Solving complex problems: Exploration and control of complex systems},
	year = {1991}}

@book{hammer1993reengineering,
	author = {Hammer, M and Champy, J},
	isbn = 9780887306402,
	pmid = 92054748,
	publisher = {Harper Business},
	title = {{Reengineering the corporation: a manifesto for business revolution}},
	year = 1993}

@article{smith1993conceptual,
	author = {Smith, G.F. and Browne, G.J.},
	importance = {2},
	journal = {Systems, Man and Cybernetics, IEEE Transactions on},
	number = {5},
	pages = {1209--1219},
	publisher = {IEEE},
	title = {Conceptual foundations of design problem solving},
	volume = {23},
	year = {1993}}

@book{goetsch1995implementing,
	author = {Goetsch, D L and Davis, S},
	isbn = 9780131808867,
	publisher = {Prentice Hall PTR},
	title = {{Implementing Total Quality}},
	year = 1995}

@article{van-de-ven1995explaining,
	author = {Van de Ven, Andrew H and Poole, Marshall Scott and others},
	journal = {Academy of management review},
	number = 3,
	pages = {510--540},
	publisher = {Academy of Management},
	title = {Explaining development and change in organizations},
	volume = 20,
	year = 1995}

@article{moffett1996model,
	author = {Moffett, Jonathan D. and Hall, Jon G. and Coombes, Andrew and McDermid, John A.},
	journal = {Journal of Requirements Engineering},
	number = {1},
	pages = {27--46},
	title = {{A Model for a Causal Logic for Requirements Engineering}},
	volume = {1},
	year = {1996}}

@article{nance1996investigation,
	address = {New York, NY, USA},
	author = {Nance, William D.},
	booktitle = {Proceedings of the 1996 ACM SIGCPR/SIGMIS Conference on Computer Personnel Research},
	doi = {10.1145/238857.238867},
	isbn = {0897917820},
	numpages = 9,
	pages = {49--57},
	publisher = {Association for Computing Machinery},
	series = {SIGCPR'96},
	title = {An Investigation of Information Technology and the Information Systems Group as Drivers and Enablers of Organizational Change},
	url = {https://doi.org/10.1145/238857.238867},
	year = 1996}

@article{hill1998positioning,
	author = {Hill, Frances M and Collins, Lee K},
	journal = {The TQM magazine},
	publisher = {MCB UP Ltd},
	title = {The positioning of {BPR} and {TQM} in long-term organisational change strategies},
	year = 1998}

@book{grobman1999improving,
	author = {Grobman, Gary M.},
	publisher = {White Hat Communications},
	title = {Improving quality and performance in your non-profit organization},
	year = 1999}

@article{valiris1999critical,
	author = {Valiris, George and Glykas, Michalis},
	journal = {Business Process Management Journal},
	number = 1,
	pages = {65--86},
	publisher = {MCB UP Ltd},
	title = {Critical review of existing {BPR} methodologies: the need for a holistic approach},
	volume = 5,
	year = 1999}

@article{jonassen2000towards,
	author = {Jonassen, David H},
	importance = {3},
	journal = {Educational technology research and development},
	number = {4},
	pages = {63--85},
	publisher = {Springer},
	title = {Towards a design theory of problem solving},
	volume = {48},
	year = {2000}}

@article{kleiner2000revisiting,
	author = {Kleiner, Art},
	journal = {Strategy+ Business},
	number = 3,
	pages = {27--31},
	title = {Revisiting reengineering},
	volume = 20,
	year = 2000}

@article{moran2000leading,
	author = {Moran, John W and Brightman, Baird K},
	journal = {Journal of Workplace Learning},
	number = 2,
	pages = {66--74},
	publisher = {MCB UP Ltd},
	title = {Leading organizational change},
	volume = 12,
	year = 2000}

@article{allen2001applying,
	author = {Allen, Richard S and Montgomery, Kendyl A},
	journal = {Organizational Dynamics},
	number = 2,
	pages = {149--161},
	publisher = {Elsevier Science},
	title = {Applying an organizational development approach to creating diversity},
	volume = 30,
	year = 2001}

@book{jackson2001problem,
	author = {Jackson, M.},
	publisher = {Addison-Wesley Publishing Company},
	title = {Problem Frames: Analyzing and Structuring Software Development Problems},
	year = {2001}}

@article{kezar2001understanding,
	author = {Kezar, A},
	doi = {10.1002/aehe.2804/abstract},
	journal = {ASHE-ERIC Higher Education Report},
	publisher = {John Wiley \& Sons, Inc.},
	title = {{Understanding and facilitating organizational change in the 21st century}},
	year = 2001}

@book{pawlowsky2001treatment,
	author = {Pawlowsky, Peter},
	journal = {Handbook of organizational learning and knowledge},
	pages = {61--88},
	publisher = {Oxford University Press New York},
	title = {The treatment of organizational learning in management science},
	year = 2001}

@article{cao2004need,
	author = {Cao, Guangming and Clarke, Steve and Lehaney, Brian},
	journal = {Systemic Practice and Action Research},
	number = 2,
	pages = {103--126},
	publisher = {Springer},
	title = {The need for a systemic approach to change management---a case study},
	volume = 17,
	year = 2004}

@article{hevner2004design,
	author = {Hevner, Alan R. and March, Salvatore T. and Park, Jinsoo and Ram, Sudha},
	doi = {10.2307/25148625},
	issn = {0276-7783},
	journal = {MIS Quarterly},
	number = {1},
	pages = {75--105},
	publisher = {Management Information Systems Research Center, University of Minnesota},
	title = {Design {Science} in {Information} {Systems} {Research}},
	url = {http://www.jstor.org/stable/25148625},
	urldate = {2021-08-20},
	volume = {28},
	year = {2004}}

@techreport{popper2004system,
	author = {Popper, Steven W and Bankes, Steven C and Callaway, Robert and DeLaurentis, Daniel},
	institution = {Potomac Institute for Policy Studies, Arlington, VA},
	number = {320},
	title = {System of systems symposium: Report on a summer conversation},
	year = {2004}}

@book{verkerk2004trust,
	author = {Verkerk, Maarten Johannes},
	publisher = {Eburon Uitgeverij BV},
	title = {Trust and Power on the Shop Floor: An Ethnographical, Ethical and Philosophical Study on Responsible Behaviour in Industrial Organisations},
	year = 2004}

@article{reijers2005best,
	author = {Reijers, Hajo A and Mansar, S Liman},
	journal = {Omega},
	number = 4,
	pages = {283--306},
	publisher = {Elsevier},
	title = {Best practices in business process redesign: an overview and qualitative evaluation of successful redesign heuristics},
	volume = 33,
	year = 2005}

@book{demers2007organizational,
	author = {Demers, C},
	isbn = 9780761929321,
	pmid = 2007004816,
	publisher = {SAGE Publications},
	title = {{Organizational Change Theories: A Synthesis}},
	year = 2007}

@inproceedings{hall2007arguing,
	author = {Hall, Jon G and Mannering, Derek and Rapanotti, Lucia},
	booktitle = {10th IEEE High Assurance Systems Engineering Symposium (HASE'07)},
	organization = {IEEE},
	pages = {23--32},
	title = {Arguing safety with problem oriented software engineering},
	year = 2007}

@article{peffers2007design,
	author = {Peffers, Ken and Tuunanen, Tuure and Rothenberger, Marcus A and Chatterjee, Samir},
	journal = {Journal of management information systems},
	number = {3},
	oro = {-},
	pages = {45--77},
	publisher = {Taylor \& Francis},
	title = {A design science research methodology for information systems research},
	volume = {24},
	year = {2007}}

@article{buchanan2008introduction,
	author = {Buchanan, Richard},
	journal = {Design Issues},
	pages = {2--9},
	publisher = {JSTOR},
	title = {Introduction: Design and organizational change},
	year = 2008}

@article{hall2008problem,
	author = {Hall, Jon~G. and Rapanotti, Lucia and Jackson, Michael},
	journal = {IEEE Transactions on Software Engineering},
	number = 2,
	pages = {226--241},
	publisher = {IEEE},
	title = {Problem oriented software engineering: Solving the package router control problem},
	volume = 34,
	year = 2008}

@incollection{netjes2009bpr,
	author = {Netjes, Mariska and Mans, Ronny S and Reijers, Hajo A and van der Aalst, Wil MP and Vanwersch, Rob JB},
	booktitle = {International Conference on Business Process Management},
	pages = {605--616},
	publisher = {Springer},
	title = {{BPR} best practices for the healthcare domain},
	year = 2009}

@incollection{barroero2010business,
	author = {Barroero, Thiago and Motta, Gianmario and Pignatelli, Giovanni},
	booktitle = {IFIP International Conference on Enterprise Architecture, Integration and Interoperability},
	pages = {32--43},
	publisher = {Springer},
	title = {Business capabilities centric enterprise architecture},
	year = 2010}

@article{lunenburg2010organizational,
	author = {Lunenburg, Fred C},
	journal = {International Journal of Management, Business, and Administration},
	number = 1,
	pages = {1--9},
	title = {Organizational development: Implementing planned change},
	volume = 13,
	year = 2010}

@incollection{mezzanotte2010applying,
	author = {{Mezzanotte Sr}, Dominic M and Dehlinger, Josh and Chakraborty, Suranjan},
	booktitle = {2010 IEEE/ACIS 9th International Conference on Computer and Information Science},
	pages = {859--863},
	publisher = {IEEE},
	title = {On applying the theory of structuration in enterprise architecture design},
	year = 2010}

@article{fischer2011process,
	author = {Fischer, Andreas and Greiff, Samuel and Funke, Joachim},
	journal = {Journal of Problem Solving},
	number = {1},
	pages = {19--42},
	title = {The process of solving complex problems},
	volume = {4},
	year = {2011}}

@inproceedings{parashar2011change,
	author = {Parashar, Prem and Bhatia, Rajesh and Kalia, Arvind},
	booktitle = {International Conference on Information Intelligence, Systems, Technology and Management},
	organization = {Springer},
	pages = {160--169},
	title = {Change impact analysis: A tool for effective regression testing},
	year = 2011}

@article{hall2012software,
	author = {Hall, Jon G and Rapanotti, Lucia},
	journal = {Innovations in Systems and Software Engineering},
	number = 3,
	pages = {175--193},
	publisher = {Springer},
	title = {Software engineering as the design theoretic transformation of software problems},
	volume = 8,
	year = 2012}

@phdthesis{uluturk2012assessment,
	address = {University of Baltimore Baltimore, Maryland},
	author = {Uluturk, Bulent},
	month = {05},
	school = {University of Baltimore School of Public Affairs, Baltimore, Maryland},
	title = {An assessment of law enforcement officers' attitudes toward Compstat model of police management},
	year = 2012}

@article{montgomerie2013owning,
	author = {Montgomerie, Johnna and Roscoe, Samuel},
	doi = {https://doi.org/10.1016/j.accfor.2013.06.003},
	issn = {0155-9982},
	journal = {Accounting Forum},
	note = {The Apple Business Model: Value Capture and Dysfunctional Economic and Social Consequences},
	number = 4,
	pages = {290--299},
	title = {Owning the consumer---Getting to the core of the {Apple} business model},
	url = {https://www.sciencedirect.com/science/article/pii/S015599821300032X},
	volume = 37,
	year = 2013}

@article{moreau2013disruptive,
	author = {Moreau, Fran{\c{c}}ois},
	journal = {International Journal of Arts Management},
	number = 2,
	title = {The disruptive nature of digitization: the case of the recorded music industry.},
	volume = 15,
	year = 2013}

@article{nkwocha2013design,
	author = {Nkwocha, A. and Hall, Jon G. and Rapanotti, L.},
	journal = {Journal of Software and Systems Modeling},
	note = {Online first (\texttt{http://www.\allowbreak{}springerlink.com/content/d45x17g438833069/})},
	number = {4},
	oro = {1},
	pages = {825-845},
	publisher = {Springer},
	title = {Design rationale capture for process improvement in the globalized enterprise: an industrial study},
	url = {https://link.springer.com/article/10.1007/s10270-011-0223-y},
	volume = {12},
	year = {2013}}

@incollection{oat2013analysis,
	author = {Oat, Elena},
	booktitle = {Seminar on Internet Working},
	publisher = {Aalto University T-110.5191},
	title = {Analysis of Netflix architecture and business model},
	year = 2013}

@article{ryan2013leading,
	author = {Ryan, Lindy},
	journal = {International Journal of Business Innovation and Research},
	number = 4,
	pages = {429--445},
	title = {{Leading change through creative destruction: how Netflix's self-destruction strategy created its own market}},
	url = {https://ideas.repec.org/a/ids/ijbire/v7y2013i4p429-445.html},
	volume = 7,
	year = 2013}

@article{lim2014multidimensional,
	author = {Lim, Wen Shien and Yazdanifard, Rashad},
	journal = {Global Perspectives on Engineering Management},
	number = 2,
	pages = {27--33},
	title = {{A Multidimensional Review on Organizational Change's Perspectives, Theories, Models, and Types of Change: Factors Leading to Success or Failure Organizational Change}},
	volume = 3,
	year = 2014}

@article{marouni2014certified,
	author = {Marouni, Herzl},
	journal = {Quality Progress},
	number = 6,
	pages = 68,
	publisher = {American Society for Quality},
	title = {The Certified Manager of Quality/Organizational Excellence Handbook},
	volume = 47,
	year = 2014}

@article{halal2015business,
	author = {Halal, William E},
	journal = {Journal of the Knowledge Economy},
	number = 1,
	pages = {31--47},
	publisher = {Springer},
	title = {Business strategy for the technology revolution: competing at the edge of creative destruction},
	volume = 6,
	year = 2015}

@article{nielsen2015systems,
	author = {Nielsen, Claus Ballegaard and Larsen, Peter Gorm and Fitzgerald, John and Woodcock, Jim and Peleska, Jan},
	journal = {ACM Computing Surveys (CSUR)},
	number = 2,
	pages = {1--41},
	publisher = {ACM New York, NY, USA},
	title = {Systems of systems engineering: basic concepts, model-based techniques, and research directions},
	volume = 48,
	year = 2015}

@inproceedings{serebrenik2015challenges,
	author = {Serebrenik, Alexander and Mens, Tom},
	booktitle = {Proceedings of the 2015 European Conference on Software Architecture Workshops},
	pages = {1--6},
	title = {Challenges in software ecosystems research},
	year = 2015}

@article{kotusev2016critical,
	author = {Kotusev, Svyatoslav},
	journal = {British Computer Society (BCS), URL: http://www. bcs. org/content/conWebDoc/55892},
	title = {The critical scrutiny of {TOGAF}},
	year = 2016}

@article{leung2017titus,
	author = {Leung, Andrew and Spyker, Andrew and Bozarth, Tim},
	journal = {Queue},
	number = 5,
	pages = {53--77},
	publisher = {ACM New York, NY, USA},
	title = {Titus: Introducing Containers to the {Netflix Cloud}: Approaching container adoption in an already cloud-native infrastructure},
	volume = 15,
	year = 2017}

@article{knauss2018continuous,
	author = {Knauss, Eric and Yussuf, Aminah and Blincoe, Kelly and Damian, Daniela and Knauss, Alessia},
	journal = {Requirements Engineering},
	number = 1,
	pages = {97--117},
	publisher = {Springer},
	title = {Continuous clarification and emergent requirements flows in open-commercial software ecosystems},
	volume = 23,
	year = 2018}

@incollection{olson2018total,
	author = {Olson, Eric W},
	booktitle = {Proceedings of the International Annual Conference of the American Society for Engineering Management.},
	pages = {1--7},
	publisher = {American Society for Engineering Management (ASEM)},
	title = {{Total Quality Management \& Apple success}},
	year = 2018}

@article{jansen2019managing,
	author = {Jansen, Slinger and Cusumano, Michael and Popp, Karl Michael},
	journal = {IEEE Software},
	number = 3,
	pages = {17--21},
	publisher = {IEEE},
	title = {Managing software platforms and ecosystems},
	volume = 36,
	year = 2019}

@article{costantini2021using,
	author = {Costantini, Silvana and Hall, Jon G. and Rapanotti, Lucia},
	doi = {10.1108/IJMPB-06-2020-0187},
	issn = {1753-8378},
	journal = {International Journal of Managing Projects in Business},
	month = jan,
	number = {5},
	pages = {1135--1162},
	publisher = {Emerald Publishing Limited},
	title = {Using complexity and volatility characteristics to guide hybrid project management},
	url = {https://doi.org/10.1108/IJMPB-06-2020-0187},
	urldate = {2021-04-05},
	volume = {14},
	year = {2021}}

@webpage{gartner2023digital,
	author = {Gartner},
	institution = {Gartner},
	journal = {Gartner Glossary},
	lastchecked = {2025-01-17},
	publisher = {Gartner, Inc.},
	title = {Digital Transformation: How to Scope and Execute Strategy},
	url = {https://www.gartner.com/en/information-technology/glossary/digital-transformation},
	year = {2023}}

@article{hall2017a-design,
	author = {Hall, Jon G and Rapanotti, Lucia},
	journal = {Information and Software Technology},
	oro = {1},
	pages = {46--61},
	publisher = {Elsevier},
	title = {A design theory for software engineering},
	url = {https://www.sciencedirect.com/science/article/abs/pii/S0950584917300691},
	volume = {87},
	year = {2017}}
